\documentclass[showpacs,preprintnumbers,amsmath,amssymb,APSl,prd,nofootinbib,superscriptaddress, twocolumn]{revtex4-1}

\usepackage{dcolumn}
\usepackage{bm}
\usepackage{ifpdf}
\usepackage{hyperref}
\usepackage{bm}
\usepackage{xcolor,color,graphicx,graphics}
\usepackage[spanish,english]{babel}
\usepackage[latin1]{inputenc}
\usepackage[OT1]{fontenc}
\usepackage{latexsym,amssymb,amsmath,amsfonts}
\usepackage{makeidx}
\usepackage{epsfig,subfigure}
\usepackage{natbib}
\usepackage{epstopdf}
\usepackage{mathrsfs}
\usepackage{hyperref}
\hypersetup{colorlinks=true, linkcolor=blue, citecolor=green}
\usepackage{enumerate}
\usepackage{xcolor}

\definecolor{red}{rgb}{1,0,0}

\def\+{^\dagger}

\def\<{\leftarrow}
\def\>{\rightarrow}

\def\({\left(}
\def\){\right)}

\def\a{\alpha} \def\b{\beta} \def\g{\gamma} \def\d{\delta} 
\def\m{\mu} \def\n{\nu} \def\r{\rho} \def\s{\sigma}  
\def\k{\kappa}\def\S{\Sigma}\def\t{\tau}

\newcommand{\bi}{\begin{itemize}} 				\newcommand{\ei}{\end{itemize}}
\newcommand{\benu}{\begin{enumerate}} 		\newcommand{\enu}{\end{enumerate}}
\newcommand{\bd}{\begin{dinglist}{0}}     \newcommand{\ed}{\end{dinglist}}
\newcommand{\bfig}{\begin{figure}[htbp]}  \newcommand{\efig}{\end{figure}}
        			
\newcommand{\bc}{\begin{center}} 				  \newcommand{\ec}{\end{center}}
\newcommand{\be}{\begin{equation}} 				\newcommand{\ee}{\end{equation}}
\newcommand{\bsub}{\begin{subequations}}  \newcommand{\esub}{\end{subequations}}
\newcommand{\ben}{\begin{eqnarray}} 			\newcommand{\een}{\end{eqnarray}}
\newcommand{\ba}[1]{\begin{array}{#1}} 		\newcommand{\ea}{\end{array}}
\newcommand{\bea}{\begin{equation}\begin{array}{rcl}}
\newcommand{\eea}{\end{array}\end{equation}}

\begin{document}
\title{Double shadows of reflection-asymmetric wormholes supported by positive energy thin-shells}


\author{Merce Guerrero} \email{merguerr@ucm.es}
\affiliation{Departamento de F\'isica Te\'orica and IPARCOS,
	Universidad Complutense de Madrid, E-28040 Madrid, Spain}
	\author{Gonzalo J. Olmo} \email{gonzalo.olmo@uv.es}
\affiliation{Departamento de F\'{i}sica Te\'{o}rica and IFIC, Centro Mixto Universidad de Valencia - CSIC.
Universidad de Valencia, Burjassot-46100, Valencia, Spain}
\affiliation{Departamento de F\'isica, Universidade Federal da
Para\'\i ba, 58051-900 Jo\~ao Pessoa, Para\'\i ba, Brazil}
\author{Diego Rubiera-Garcia} \email{drubiera@ucm.es}
\affiliation{Departamento de F\'isica Te\'orica and IPARCOS,
	Universidad Complutense de Madrid, E-28040 Madrid, Spain}

\date{\today}
\begin{abstract}
We consider reflection-asymmetric thin-shell wormholes within Palatini $f(\mathcal{R})$ gravity using a matching procedure of two patches of electrovacuum space-times at a hypersurface (the shell) via suitable junction conditions. The conditions for having (linearly) stable wormholes supported by positive-energy matter sources are determined. We also identify some subsets of parameters able to locate the shell radius above the event horizon (when present) but below the photon sphere (on both sides). We illustrate with an specific example that such two photon spheres allow an observer on one of the sides of the wormhole to see another (circular) shadow in addition to the one generated by its own photon sphere, which is due to the photons passing above the maximum of the effective potential on its side and bouncing back across the throat due to a higher effective potential on the other side. We finally comment on the capability of these double shadows to seek for traces of new gravitational physics beyond that described by General Relativity.

\end{abstract}

\maketitle

\section{Introduction}

We are currently witnessing the beginning of an era in which the technological leap in the capability of our observational devices, together with the progress in the development of powerful numerical methods and the establishment of large international collaborations is beginning to pay off in the testing of the strong-field regime of the gravitational interaction \cite{Amendola:2012ys,Berti:2015itd,Barack:2018yly,Barausse:2020rsu}. In the search for new physics in such a regime beyond that of Einstein's General Relativity (GR) when using compact objects, one of the main challenges is to find observational discriminators with respect to canonical GR compact objects (either black holes or stellar objects) \cite{Cardoso:2019rvt,Carballo-Rubio:2018jzw}. In this sense, the topic of modified theories of gravity has shaped a landscape made up of a large and extremely rich number of proposals to extend GR  (see e.g. \cite{DeFelice:2010aj,CLreview,Clifton:2011jh,Nojiri:2017ncd,BeltranJimenez:2017doy,Heisenberg:2018vsk,Olmo:2019flu} for some reviews). Despite the fact that  many new rotating black holes \cite{Bambi:2013ufa,Cembranos:2011sr,Ayzenberg:2014aka,Buoninfante:2018xif,Cano:2019ore,Bahamonde:2020snl} and other exotic compact objects \cite{Schunck:2003kk,Mathur:2005zp,Visser:2003ge,Herdeiro:2014goa} have been obtained in the literature, most of them introduce quantitatively small differences with respect to GR solutions, in such a way that the potential for discrimination between theories  is greatly diminished. Therefore, there is a gap for new ideas on the search for qualitatively new predictions from compact objects beyond GR.

A natural place to look for discrepancies between the predictions of GR and modified theories of gravity is by studying phenomena that occur near the photon sphere of a compact object. This is so because, being the photon sphere the last (unstable) orbit of a null geodesic around a compact object such that any small perturbation will turn any photon to spiral down into the object or to escape to infinity, it represents the closest region to the (in the case of a black hole) event horizon from which information can be recovered for asymptotic observers. Moreover, since this is a phenomenon occurring in the strong-field regime, one would expect to find there the most noticeable departures from GR expectations. In this sense, several devices are currently operating in order to look for signals related to such critical curves: the LIGO/VIRGO Collaboration for gravitational waves out of black hole mergers \cite{Abbott:2016blz}, and the Event Horizon Telescope for the shadow originated from the plasma  around black holes illuminated by electromagnetic radiation  \cite{Akiyama:2019cqa,Psaltis:2018xkc,Johnson:2019ljv}. Therefore, there is nowadays a strong interest in the theoretical characterization of black hole mimickers: ultracompact objects (perhaps having a photon sphere), almost identical to black holes but able to be distinguished from them via the absence of the event horizon \cite{Cardoso:2017cqb,Abdikamalov:2019ztb,Olivares:2018abq}.

One historical candidate for black hole mimickers is given by wormholes \cite{VisserBook}. Indeed, wormholes arise as exact solutions of Einstein's equations, and are typically interpreted as structures allowing to connect far away regions of the space-time via the restoration of geodesic completeness. The nature of these objects could be revealed both by their different shadows as compared to black holes \cite{Tsukamoto:2012xs,Shaikh:2018oul,Amir:2018pcu,Shaikh:2018kfv},  or by the fact that the absence of an event horizon is capable to trigger new phenomena, such as the generation of echoes, namely, secondary periodic gravitational waves of decreasing amplitude generated due to the modes trapped in the effective potential between the photon sphere and the wormhole throat \cite{Cardoso:2016oxy}, a feature which could be probed in the next few years \cite{Micchi:2020gqy}. However, as opposed to their black hole cousins, wormholes face the fundamental difficulty that they can only be supported by matter fields violating the energy conditions and, as such, their physical plausibility has been long contested in the literature. To ameliorate this difficulty, in thin-shell wormholes one cuts two patches of some space-times and pastes them together at some hypersurface (the shell) using a suitable junction condition formalism, and in a second step the shell is made as small as possible to keep the violations of the energy conditions confined to the tiniest possible region \cite{Visser:1989kg,Visser:2003yf}. If the radius of the shell is located above the would-be Schwarzschild radius (for spherically symmetric objects) then one gets a traversable wormhole \cite{Garcia:2011aa}.

 Traversable, thin-shell wormholes are usually constructed assuming the same space-time patch on both sides of its throat, but nothing prevents one from considering reflection-asymmetric wormholes, namely, different solutions on each side \cite{Forghani:2018gza,Wang:2020emr}. For the sake of this work we are interested in the results of Ref. \cite{Wielgus:2020uqz}, where the authors construct reflection-asymmetric wormholes from two patches of Reissner-Nordstr\"om (RN) space-times of different masses and charges. They show that, in addition to the standard shadow on each side, on the side with the lower peak in the effective potential another shadow is present, which is originated from those photons whose impact factor allows them to overcome the maximum of the effective potential and bounce back across the wormhole throat after hitting the potential slope on the other side (for an stability analysis of these solutions see \cite{Tsukamoto:2021fpp}). This phenomenon, known as double shadows, could hint at the existence of new physics beyond the one of GR. The construction of \cite{Wielgus:2020uqz}, despite being gravity-model-independent, still has the potential problem of the violation of the energy conditions at the shell. However, this difficulty can go away if we frame wormholes within the context of modified gravity, where the existence of additional gravitational corrections, understood as a kind of matter fluid, are able to restore geodesic completeness without unavoidably violating the energy conditions, see e.g. \cite{Eiroa:2008hv,Richarte:2007zz,Son:2010bs,Harko:2013yb,Bronnikov:2015pha,Bahamonde:2016ixz,Zubair:2017hsq,Shaikh:2018yku,Lobo:2020jfl,Singh:2020rai}.

 The main aim of this work is therefore to build reflection-asymmetric thin-shell wormholes from two patches of RN geometries with different masses and charges using a Palatini $f(\mathcal{R})$ extension of GR. In the Palatini formalism, metric and affine connection are regarded as independent entities, which in the $f(\mathcal{R})$ case yields a suitable framework to test the strong-field regime of the gravitational interaction without bumping into conflicts with current solar system observations and gravitational wave astronomy results \cite{Olmo:2011uz}. Recently, some of us \cite{Olmo:2020fri} developed  a suitable junction conditions formalism for this theory, and moreover found  spherically symmetric stable thin-shell wormholes supported by positive energy matter fields \cite{Lobo:2020vqh}. Here we shall extend these results to the asymmetric case. We shall show that while the matching of a Schwarzschild solution with a RN one does  not allow for such stable and positive-energy asymmetric wormholes, the RN-RN matching do contain such solutions for some regions in the space of parameters. Moreover, we shall show that there is a subset of them such that the shell radius is above the event horizon (when present) but below the photon sphere radius (on both sides of the shell), therefore finding the existence of  stable, positive-energy, traversable thin-shell wormholes having two distinct photon spheres. We shall illustrate our results with an explicit example and construct the corresponding light trajectories and double shadow.

\section{Theoretical framework}

\subsection{Junction conditions in Palatini $ f(\mathcal{R}) $ gravity}

Let us start our analysis by considering two smooth manifolds $ \mathcal{M}_\pm $ (with metrics $ g_{\m\n}^\pm $), bounded by some hypersurfaces $ \Sigma_\pm $. In the thin-shell formalism one matches both bulks to form a single manifold $ \mathcal{M} = \mathcal{M}_+ \cap \mathcal{M}_- $, which has a time-like hypersurface $ \Sigma=\Sigma_\pm $ where both manifolds are joined and coincide with their respective boundaries. Since there may be discontinuities on several geometric and matter quantities across the hypersurface, the appropriate mathematical framework to deal with this scenario is to upgrade the concept of tensorial functions to that of tensorial distributions. The corresponding conditions that the geometry and the matter fields need to satisfy at the matching hypersurface and across it are called {\it junction conditions} \cite{JC1,JC2}. One basic such condition is that the space-time metric components $ g_{\m\n} $ have to be continuous across the hypersurface, that is
\begin{equation}\label{continuity metric}
 g_{\m\n}^+|_{_\S} = g_{\m\n}^-|_{_\S} \ .
\end{equation}
The remaining junction conditions are heavily influenced by the theory of gravity chosen and its corresponding field equations. Therefore, we need now to define our specific framework.

The action of $ f(\mathcal{R}) $ gravity is given by
\begin{equation} \label{eq:action}
\mathcal{S}=\frac{1}{2\kappa^2} \int d^4x \sqrt{-g} f(\mathcal{R}) + \int d^4x \sqrt{-g}  \mathcal{L}_m(g_{\mu\nu},\psi_m) \ ,
\end{equation}
where $\kappa^2$ is Newton's constant in suitable units, $g$ is the determinant of the space-time metric $g_{\mu\nu}$, $ f(\mathcal{R}) $ is a function of the Ricci scalar $ \mathcal{R}\equiv g^{\m\n}\mathcal{R}_{\m\n} (\Gamma)$, which is constructed out of the
 Ricci tensor  as $\mathcal{R}_{\mu\nu}(\Gamma) \equiv {\mathcal{R}^\alpha}_{\mu\alpha\nu}(\Gamma)$, which only depends on the affine connection $\Gamma \equiv \Gamma_{\mu\nu}^{\lambda}$, the latter being independent of the metric (Palatini approach). As for the matter Lagrangian, $  \mathcal{L}_m $, it is a function of the space-time metric and the matter fields $\psi_m$, but not of the independent connection.

In the Palatini formulation, by taking the trace of the field equations associated to the variation of the action \eqref{eq:action} with respect to the metric, one finds that \cite{Olmo:2011uz}
\begin{equation}\label{eq:trace-field eq}
\mathcal{R} f_\mathcal{R} -2f=\k^2 T \ ,
\end{equation}
where $ f_\mathcal{R} \equiv d f / d\mathcal{R} $ and $ T = g^{\m\n}T_{\m\n} $ is the trace of the stress-energy tensor defined as $ T_{\m\n}= \frac{2}{\sqrt{-g}}\frac{\d \mathcal{S}_m}{\d g^{\m\n}} $. The above equation makes a clear distinction with respect to the metric formulation of $f(R)$ theories \cite{Olmo:2005zr}, in that it represents an algebraic relation between the Ricci scalar and the matter fields, $ \mathcal{R}\equiv \mathcal{R}(T) $, which implies that the curvature can be effectively removed out in terms of the matter fields (being thus closer to the spirit of GR, where $R=-\kappa^2 T$). Due to this property, and to the fact that the equation for the independent connection, $\nabla_{\mu}^{\Gamma}(\sqrt{-g} f_{\mathcal{R}} g^{\mu\nu})=0$, implies that $\Gamma$ is Levi-Civita of $q_{\mu\nu} \equiv f_{\mathcal{R}}(T) g_{\mu\nu}$, then Palatini $f(\mathcal{R})$ gravity and its generalizations are sometimes interpreted as GR with extra couplings in the matter fields \cite{Afonso:2017bxr}. For the sake of this paper, the fact that the field equations of Palatini $f(\mathcal{R})$ gravity are second-order introduces significant changes in the shape of its junction conditions not only with respect to their metric cousins \cite{Seno}, but surprisingly with GR itself.

Introducing the matter content on $\mathcal{M}_{\pm}$ as given by some energy-momentum tensors $ T_{\m\n}^\pm $, and denoting by $S_{\mu\nu}$ their restriction to the hypersurface $\Sigma$ (with $S\equiv {S^\mu}_{\mu}$ denoting its trace), the distributional version of the field equations of Palatini $f(R)$ gravity straightforwardly leads to the conditions \cite{Olmo:2020fri}
\begin{equation}\label{eq: stress-energy cond}
[ T ] = 0 \text{ and } S= 0 \ ,
\end{equation}
where brackets denote the discontinuity of the quantity inside them across $\Sigma$, that is, $ [T] = T^+|_{_\S}-T^-|_{_\S}$. Comparing such conditions to the ones in GR one can see that the second one is absent there \cite{JC1,JC2}.

There are further junction conditions. Let us define the pullback of the first fundamental form on the hypersurface as $ h_{\m\n} = g_{\m\n}-n_\m n_\n $, where $ n_\m $ is the unit vector normal to $ \S $, and the pullback of the second fundamental form (the extrinsic curvature) as
\begin{equation}\label{second fundamental form}
K^\pm_{\m\n}\equiv{ h^\r}_\m {h^\s}_\n\nabla_\r^\pm\, n_\s \ .
\end{equation}
Plugging this result into the singular part of the Palatini  field equations on the shell and applying the junction conditions found in Eq.\eqref{eq: stress-energy cond} one gets
\begin{equation}\label{eq: Einstein field eq on the shell}
-[K_{\m\n}]+\frac{1}{3}h_{\m\n}[{K^\r}_\r]=\kappa^2 \frac{S_{\m\n}}{f_{\mathcal{R}_\S}} \ ,
\end{equation}
where the subscript $\S $ means that the function is evaluated in the shell. Note that this equation departs from its GR counterpart, $ -[K_{\m\n}]+h_{\m\n}[{K^\r}_\r]  = \kappa^2 S_{\m\n}$, since the Lagrangian density of GR with a cosmological constant, $ f(\mathcal{R}) = \mathcal{R} - 2 \Lambda$, is a singular case in this formalism (which justifies why the second condition of (\ref{eq: stress-energy cond}) does not hold in GR, being replaced by $ [{K^\r}_\r]= S/2 $).

The last two junction conditions are obtained from the Bianchi identities on the shell. These can be expressed as energy conservation equations as
\begin{eqnarray}
D^{\rho}S_{\rho\nu}& =&-n^{\rho}{h^\sigma}_{\nu}[T_{\rho\sigma}] \label{Bianchi ident} \\
(K^+_{\r\s}+K^-_{\r\s})S^{\r\s}&=&2n^{\rho}n^{\sigma}[T_{\rho\sigma}] - \frac{ 3\mathcal{R}_T^2 f_{\mathcal{R}\mathcal{R}}^2}{f_{\mathcal{R}}}[b^2] \label{Bianchi ident2} \ ,
\end{eqnarray}
where $\mathcal{R}_T \equiv d\mathcal{R}/dT$ and $b \equiv n^{\mu}[\nabla_{\mu} T]$.

\subsection{Electrovacuum spherically symmetric space-times}

Let us consider two static, spherically symmetric space-times on $\mathcal{M}_{\pm}$ defined as
\begin{equation}
ds^2=A_{\pm}(r_{\pm})dt^2-B^{-1}_{\pm}(r_{\pm})dr^2-r_{\pm}^2 d\Omega^2 \ ,
\end{equation}
where $A_{\pm}(r_{\pm}),B_{\pm}(r_{\pm})$ are the metric functions on each side and $d\Omega^2=d\theta^2 + \sin^2 \theta d\varphi^2$ is the unit volume element on the two-spheres. The induced metric on the shell is written as
\begin{equation}
ds_{\Sigma}^2=-d \tau^2 + R^2(\tau)d\Omega^2 \ ,
\end{equation}
and is parameterized in terms of the proper time $\tau$ of a comoving observer in $\Sigma$, with $R(\tau)$ representing its areal radius.

The non-vanishing components of the second fundamental form $ {K^i}_j=\text{diag}\,({K^\tau}_\t,\,{K^\theta}_\theta,\,{K^\theta}_\theta) $ in Eq.(\ref{second fundamental form}) have been computed in \cite{Lobo:2020vqh} and are given by the expressions
\begin{eqnarray}
{K^\tau}_{\tau_\pm}&=&\pm\frac{R_\tau ^2 \left(A_{R\pm} B_\pm-A_\pm B_{R\pm}\right)+2 A_\pm B_\pm R_{\tau \tau} +A_{R\pm} B_\pm^2}{2 A_\pm B_\pm \sqrt{B_\pm+R_\tau ^2}}\label{1} \nonumber\\
{K^{\theta }}_{\theta _\pm}&=&\pm\frac{\sqrt{B_\pm+R_\tau ^2}}{R} \ ,\label{2} \nonumber
\end{eqnarray}
where we have introduced the notation $A_R \equiv dA/dR$\footnote{We point out that while the metric components must be continuous across $\Sigma$ this does not need to be so for their derivatives, which is by we have kept the $\pm$ in labelling the metric functions and their derivatives there.}. As for the matter content of our shell we assume a perfect fluid of the form ${S^\mu}_{\nu}=\text{diag}(-\sigma,\mathcal{P},\mathcal{P})$, where $\sigma$ and $\mathcal{P}$ are the energy density and pressure, respectively. Remarkably, the second of the junction conditions (\ref{eq: stress-energy cond}) tells us that $\mathcal{P}=\sigma/2$, which means that the energy content of the shell is entirely determined by its energy density, and therefore no equation of state is required to close the system. Now, the junction condition (\ref{eq: Einstein field eq on the shell}) with this matter content  yields the equation
\begin{equation}\label{3}
\left[K^\tau _{\tau }\right]-\left[K^{\theta }_{\theta }\right] =\dfrac{3\kappa^2}{2f_{R_\Sigma}}\sigma \ .
\end{equation}
We now call upon the energy conservation equation (\ref{Bianchi ident}), which for the present spherically symmetric case reads
\begin{equation}
-D_{\rho}{S^\rho}_{\nu}=-\dot{\sigma} +\frac{2\dot{R}}{R}(\sigma+\mathcal{P}) \ .
\end{equation}
Equating this to the energy conservation equation (\ref{Bianchi ident}) one gets
\begin{equation} \label{eq:eqenergy}
-\dot{\sigma} +\frac{2\dot{R}}{R}(\sigma+\mathcal{P})=-n^{\rho}{h^\sigma}_{\nu}[T_{\rho\sigma}] \ ,
\end{equation}
which relates quantities of the fluid defined on the shell with the discontinuity in the energy-momentum tensor across the normal direction to it.

Let us now consider electrovacuum space-times. For any electrostatic, spherically symmetric field described by a given non-linear electrodynamics (defined by the two field invariants of the electromagnetic field), the corresponding energy-momentum tensor can be written as ${T_\mu}^{\nu}=\text{diag}(-\phi_1(r),-\phi_1(r),\phi_2(r),\phi_2(r))$, where the functions $\{\phi_1,\phi_2\}$ characterize each particular model (for instance, in Maxwell electrodynamics, $\phi_1=\phi_2=-q^2/r^4$).  In such a case, one can show  that the quantities in the bracket of the right-hand side of Eq.(\ref{eq:eqenergy}) vanish on both sides of $\Sigma_{\pm}$ \cite{Lobo:2020vqh} and, therefore, also does its discontinuity across $\Sigma$. Under such conditions, and recalling that $\mathcal{P}=\sigma/2$, then Eq.(\ref{eq:eqenergy}) integrates as
\begin{equation} \label{eq:sigmaR}
\sigma=\frac{C}{R^3} \ ,
\end{equation}
(where $C$ is an integration constant), which nicely relates the energy density of the matter fields with the radius of the shell. This completes our formalism for the analysis of reflection-asymmetric thin-shell wormhole solutions. In the next two sections we shall seek explicit examples of this construction using electrovacuum space-times.

\section{Traversable wormholes from surgically joined Schwarzschild-RN space-times}

We shall first start by considering the simplest  case leading to reflection-asymmetric solutions, as given by the matching of a Schwarzschild and a RN space-time. Our aim here is to study whether the resulting thin-shell wormholes can be supported by matter energy sources on the shell with a positive energy density, and if such a construction is stable against small perturbations. We first note that for any $f(\mathcal{R})$ theory, the fact that the dependence on the curvature goes via the trace of energy-momentum tensor, $\mathcal{R}=\mathcal{R}(T)$, implies that in vacuum (Schwarzschild) or for a vanishing trace (Maxwell), one has that $\kappa^2/f_{\mathcal{R}_{\Sigma}}=\tilde{\kappa}^2=$constant, regardless of the form of the $f(\mathcal{R})$ Lagrangian chosen. Moreover, thanks to this property both the first of the junction conditions (\ref{eq: stress-energy cond}) and (\ref{Bianchi ident}) automatically hold.

If we turn now our attention to Eq.(\ref{3}), recalling the above property, and combining with (\ref{eq:sigmaR}), one gets
\begin{equation} \label{eq:matcheq}
({K^\tau}_{\tau}^{+} - {K^\theta}_{\theta}^{+}) + ({K^\tau}_{\tau}^{-} - {K^\theta}_{\theta}^{-})=\frac{3\tilde{\kappa}^2}{2} \sigma \ .
\end{equation}
Now, let us consider a Schwarzschild space-time in $ \mathcal{M}_+ $ and a RN one in $ \mathcal{M}_- $, that is
\begin{eqnarray}
ds_{+}^2&=&-\Big(1-\frac{2M_+}{r}\Big)dt^2+\frac{dr^2}{1-\frac{2M_+}{r}} + r^2 d\Omega^2 \nonumber \\
ds_{-}^2&=&-\Big(1-\frac{2M_-}{r}+\frac{Q_-^2}{r^2}\Big)dt^2+\frac{dr^2}{1-\frac{2M_-}{r}+\frac{Q_-^2}{r^2}} + r^2 d\Omega^2 \nonumber
\end{eqnarray}
where $(M_+,M_-)$ are the corresponding masses and $Q_-$ the charge. This turns Eq.(\ref{eq:matcheq}) into
\begin{eqnarray} \label{Comb second fund form}
\dfrac{\gamma}{R^3 }&=& \frac{3 M_++R \left(R R_{\tau \tau}-R_\tau^2-1\right)}{R^2 \sqrt{1-\frac{2 M_+}{R}+R_\tau^2}} \nonumber \\
&+&\frac{3 M_- R-2 Q_-^2+R^2 \left(R R_{\tau \tau}-R_\tau^2-1\right)}{R^3 \sqrt{1-\frac{2 M_-}{R}+\frac{Q_-^2}{R^2}+R_\tau^2}} \ ,
\end{eqnarray}
where $ \gamma=\tfrac{3\tilde{\kappa} C}{2}$ and we recall that $R$ refers to the shell radius these two space-times are matched at. Note that the continuity on the metric components across $\Sigma$, Eq.(\ref{continuity metric}), for the above two line elements gives the extra condition
\begin{equation} \label{eq:dondn}
Q_-^2=2R(M_--M_+) \ ,
\end{equation}
which requires the masses on each side to be different in order to have a non zero amount of charge (otherwise one would be matching two Schwarzschild space-times, yielding a symmetric wormhole). Obviously, for constant $M_\pm$ and $Q_-$, this condition can only be met in those static configurations for which there exist stable solutions with $R_0=Q_-^2/2(M_--M_+)$.

In order to study the (linear) stability of these solutions, we first note that Eq.(\ref{Comb second fund form}) can be rewritten in a more convenient form as
\begin{equation}\label{R tau tau Schw}
R_{\tau \tau}=\frac{ \gamma - \frac{3 M_+ R-R^2 \left(R_\tau^2+1\right)}{\sqrt{1-\frac{2 M_+}{R}+R_\tau^2}} -\frac{3 M_- R-2 Q_-^2-R^2 \left(R_\tau^2+1\right)}{\sqrt{1-\frac{2 M_-}{R}+\frac{Q_-^2}{R^2}+R_{\tau }^2}}}{R^3 \Bigg( \frac{1}{ \sqrt{1-\frac{2 M_-}{R}+\frac{Q_-^2}{R^2}+R_{\tau }^2}}+\frac{1}{\sqrt{1-\frac{2 M_+}{R}+R_\tau ^2}}\Bigg) } \ .
\end{equation}
We shall now assume that there is an equilibrium configuration given by $R_\tau=0$. Therefore, by expanding Eq.\eqref{R tau tau Schw} in series of $ (R-R_0) $, where $ R_0 $ is the shell radius of that equilibrium configuration, one finds, to first order:
\begin{equation}\label{expansion R tau tau}
R_{\tau \tau} \approx C_1(R_0) +C_2(R_0) (R-R_0) + \mathcal{O}(R-R_0)^2,
\end{equation}
where $ C_1 $ and $ C_2 $ are some cumbersome functions of the shell radius $ R_0 $, the parameter $ \gamma $, the masses  of each side, and the charge. In order to have an equilibrium configuration the first term in this expression must be zero, $C_1=0$, while the second must be negative for such an equilibrium to be stable. This first condition allows us to solve for the parameter $\gamma$ as
\begin{equation}\label{gamma Schw}
\gamma_{_{SRN}}= -\frac{R_0^2-3 M_- R_0+2 Q_-^2}{\sqrt{1-\tfrac{2 M_-}{R_0}+\tfrac{Q_-^2}{R_0^2}}}-\frac{R_0^2-3 M_+ R_0}{\sqrt{1-\tfrac{2 M_+}{R_0}}} \ .
\end{equation}
Therefore, this expression links the energy density to the masses and charge of the solutions as well as to the radius of the shell. Since we are interested in the equilibrium configurations supported by positive energy sources, $ \gamma_{_{SRN}}>0 $, we must deal with this expression with care in order to sort its dependence on its variables. To do it so,  we find it convenient to define the following dimensionless variables
\begin{eqnarray}
R&= x \, M_- \label{R dim}\\
M_+ &= \xi \, M_- \label{M ratio}\\
Q^2_-&= y \,M^2_-, \label{Q dim}
\end{eqnarray}
In terms of these variables, the continuity of the metric parameters across the shell, Eq.(\ref{eq:dondn}), becomes simply
\begin{equation}\label{key}
\xi=1-\dfrac{y}{2x_0} \ ,
\end{equation}
so that the parameter  $\gamma$ in (\ref{gamma Schw}) turns into
\begin{eqnarray}
\tilde{\gamma}_{_{SRN}}& \equiv &\frac{\gamma_{_{SRN}}}{M_-^2}=-x_0\frac{4 (x_0-3) x_0+7 y}{2 \sqrt{x_0(x_0-2)+y}} \ . \label{gammaSRN}
\end{eqnarray}
From the above equation it is straightforward to find the conditions for its positivity. Indeed, these amount to the fulfilment of
\begin{equation} \label{eq:srncon1}
x_0>\frac{2}{3}; \quad x_0(2-x_0)<y<\frac{4x_0(3-x_0)}{7} \ ,
\end{equation}
where the second condition also enforces that $x_0<3$.

\begin{figure}[t!]
	\centering
	\includegraphics[width=0.45\textwidth]{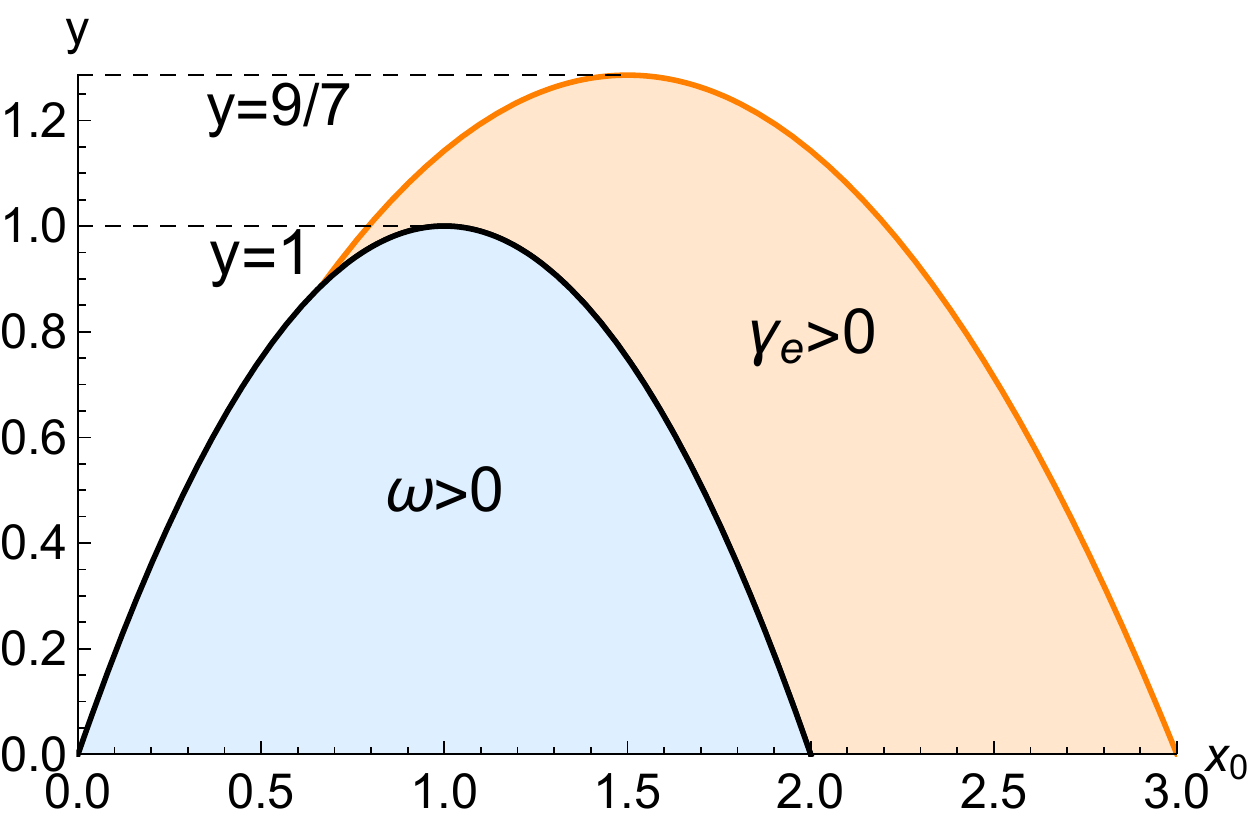}
	\caption{Parameter space of $ \tilde{\gamma}_{_{SRN}}$ in Eq.(\ref{gammaSRN}) and of $ \omega_{_{SRN}}^2>0 $ in Eq.(\ref{omegaSRN}) as a function of the dimensionless radius of the shell, $ x_0 $, and of the charge-to-mass ratio $y=Q_-^2/M_-^2$. The orange curve corresponds to $ y= (4x_0(3-x_0))/7 $, which is the upper limit of the (orange-shaded) region where the energy density is positive (check Eq.(\ref{eq:srncon1})). On this boundary curve, we have $ \tilde{\gamma}_{_{SRN}}=0$. On the other hand, the black curve is given by $y= x_0(2-x_0)$. The portion in contact with the orange shaded region, from $2/3<x_0<2$,  represents both the lower limit of $ \tilde{\gamma}_{_{SRN}}>0 $ and the upper limit of the (blue-shaded) region $ \omega^2_{_{SRN}}>0 $. On this curve, we have $ \tilde{\gamma}_{_{SRN}}\to +\infty$ and $ \omega^2_{_{SRN}}\to \pm \infty$, with the plus sign in the blue region. Given that the transition from the orange region to the blue one is not continuous, one concludes that  there is no intersection between them, meaning that there are no stable positive-energy configurations. Note also that traversable wormholes could only exist in the region $x_0>2$ in order to avoid the presence of horizons in both the Schwarzschild and RN sides, but this region has no stable solutions.}
	\label{fig:stabilitysrn}
\end{figure}

Let us now deal with the stability of these solutions. Going back to the perturbation equation, (\ref{expansion R tau tau}), and replacing the value of $\tilde{\gamma}_{ SRN}$ of (\ref{gammaSRN}) there, one finds an expression of the form
\begin{equation}\label{perturbation equation}
\delta_{t t} + \omega^2_{_{SRN}}\delta(t)=0,
\end{equation}
where $ \delta\equiv R-R_0 $, and we have introduced the dimensionless time variable $ \t \rightarrow t \, M_-  $, while the parameter
\begin{equation}\label{omegaSRN}
\omega_{_{SRN}}^2= \frac{-8 \left(2 x_0^2-8 x_0+9\right) x_0^2+4 (17-7 x_0) x_0\, y-17 y^2}{8 x_0^4 \left((x_0-2) x_0+y\right)}  \ .
\end{equation}
In order to have perturbations of bounded amplitude leading to stable solutions we have to require that $ \omega_{_{SRN}}^2 >0 $. This yields the conditions
\begin{equation} \label{eq:srncon2}
x_0>0; \quad  y<x_0(2- x_0) \ .
\end{equation}
For the purpose of having stable thin-shell wormholes supported by positive energy-density matter sources both sets of conditions (\ref{eq:srncon1}) and (\ref{eq:srncon2}) must hold at the same time. However, it is readily seen that this cannot happen, since the curve $y=x_0(2-x_0)$ bounds the region (from below and from above, respectively) in which each condition is satisfied and, therefore, there is no overlap between them (see Fig. \ref{fig:stabilitysrn}). Thus, this type of solutions cannot arise by matching a Schwarzschild and RN space-times for any values of the masses and charge of the solutions. In the next section we shall generalize this framework to deal with two RN solutions on each side.

\section{Traversable wormholes from surgically joined RN space-times}

Let us now assume two RN space-times, described by their respective masses and charges, that is
\begin{eqnarray*}
ds_{+}^2&=&-\Big(1-\frac{2M_+}{r}+\frac{Q_+^2}{r^2}\Big)dt^2+\frac{dr^2}{1-\frac{2M_+}{r}+\frac{Q_+^2}{r^2}} + r^2 d\Omega^2  \\
ds_{-}^2&=&-\Big(1-\frac{2M_-}{r}+\frac{Q_-^2}{r^2}\Big)dt^2+\frac{dr^2}{1-\frac{2M_-}{r}+\frac{Q_-^2}{r^2}} + r^2 d\Omega^2
\end{eqnarray*}
We shall now follow the same procedure as in the previous section. Then Eq.(\ref{R tau tau Schw}) generalizes to
\begin{equation}\label{R tau tau RNRN}
R_{\tau \tau}=\frac{ \gamma - \frac{3 M_+ R -2Q_+^2-R^2 \left(R_\tau^2+1\right)}{\sqrt{1-\frac{2 M_+}{R} +\frac{Q_+^2}{R^2}+R_\tau^2}} -\frac{3 M_- R-2 Q_-^2-R^2 \left(R_\tau^2+1\right)}{\sqrt{1-\frac{2 M_-}{R}+\frac{Q_-^2}{R^2}+R_{\tau }^2}}}{R^3 \Bigg( \frac{1}{ \sqrt{1-\frac{2 M_-}{R}+\frac{Q_-^2}{R^2}+R_{\tau }^2}}+\frac{1}{\sqrt{1-\frac{2 M_+}{R}+\frac{Q_+^2}{R^2}+R_\tau ^2}}\Bigg) }
\end{equation}
with the same definitions as before. Expanding in series of $(R-R_0$) and demanding the zeroth term in such an expansion to vanish, one generalizes (\ref{gamma Schw}) to
\begin{equation}\label{gammaRNRN}
\gamma_{_{RN}}= -\frac{R_0^2-3 M_- R_0+2 Q_-^2}{\sqrt{\frac{R_0^2-2 M_- R_0+Q_-^2}{R_0^2}}}-\frac{R_0^2-3 M_+ R_0+2 Q_+^2}{\sqrt{\frac{R_0^2-2 M_+ R_0+Q_+^2}{R_0^2}}} \ .
\end{equation}
Our next step is to rewrite the expression above under a more compact form. To do it,  we use again the dimensionless variables introduced in Eqs.\eqref{R dim}-\eqref{Q dim}, with the addition of
\begin{equation}\label{Q ratio}
Q_+^2 = \eta \, Q_-^2 \ .
\end{equation}
This way, the condition on the continuity of the metrics \eqref{key} generalizes to
\begin{equation}\label{xi relation}
\xi = 1- \dfrac{y}{2 x_0}(1-\eta)  \ ,
\end{equation}
while Eq.\eqref{gammaRNRN} becomes
\begin{equation}\label{gamma RN}
\tilde{ \gamma}_{_{RN}}=-x_0 \frac{4 (x_0-3) x_0+(\eta +7) y}{2 \sqrt{(x_0-2) x_0+y}} \ ,
\end{equation}
which recovers the expression (\ref{gammaSRN}) of the previous case when $\eta=0$ (i.e., when $Q_+=0$).  The above expression allows us to classify two regions for its positiveness
\begin{eqnarray}
&&2/3<x_0< 2 \quad \text{and} \quad  0<\eta <\frac{3 x_0-2}{2-x_0} \\
&&x_0\geq 2 \quad \quad \quad \quad \text{and} \quad \eta>0
\end{eqnarray}
supplemented with the additional condition
\begin{equation} \label{eq:acrn}
x_0(2-x_0)<y<\frac{4x_0(3-x_0)}{\eta +7} \ .
\end{equation}
This region is qualitatively similar to that shown in orange in Fig. \ref{fig:stabilitysrn}, which depicts the $\eta=0$ case. This last condition can actually be split into two: if $0<y<1$ then the radius of the shell must satisfy $x_0>1+\sqrt{1-y}$, while if $1<y<\tfrac{9}{7+\eta}$ then it must be contained in the interval
\begin{equation}
\frac{3-\sqrt{9-(7+\eta)}}{2}<x_0<\frac{3+\sqrt{9-(7+\eta)}}{2} \ .
\end{equation}

Substituting now the expression of $ \tilde{\gamma}_{_{RN}} $ of Eq.\eqref{gamma RN} into the expansion of $ R_{\t\t} $, one gets formally the same equation as in \eqref{perturbation equation} but in this case $ \omega_{_{RN}} ^2$ is given by
\begin{eqnarray}\label{omega RN}
\omega_{_{RN}}^2&=&-\frac{8 \left(2 x_0^2-8 x_0+9\right) x_0^2+4 x_0 y (\eta -(\eta -7) x_0-17)}{8 x_0^4 \left((x_0-2) x_0+y\right)} \nonumber \\
&-&\dfrac{\left((\eta-1) ^2+16\right) y^2}{8 x_0^4 \left((x_0-2) x_0+y\right)} \ .
\end{eqnarray}
There are again two regions in which the condition  $ \omega_{_{RN}}^2>0 $ for stability is fulfilled:
\begin{eqnarray}
&& y <x_0(2-x_0) \label{omega RN1} \\
&& \dfrac{a-2\sqrt{b}}{(\eta-1) ^2+16}<y<\dfrac{a+2\sqrt{b}}{(\eta-1) ^2+16} \ , \label{omega RN2}
\end{eqnarray}
where in the last condition we have defined the constants
\begin{eqnarray}
	a&=&2  x_0\left(\eta  (x_0-1) +17-7 x_0   \right)\\
	b&=&-3 \eta ^2 x_0^4-6 \eta  x_0^4-19 x_0^4+14 \eta ^2 x_0^3+16 \eta  x_0^3 \nonumber  \\
	&&+34 x_0^3-17 \eta ^2 x_0^2+2 \eta  x_0^2-17 x_0 \ .
\end{eqnarray}
Note that the region (\ref{omega RN1}) can also be spelled out as $1-\sqrt{1-y}<x_0<1+\sqrt{1-y}$ provided that $0<y<1$.

\begin{figure}[t!]
	\centering
	\includegraphics[width=0.44\textwidth]{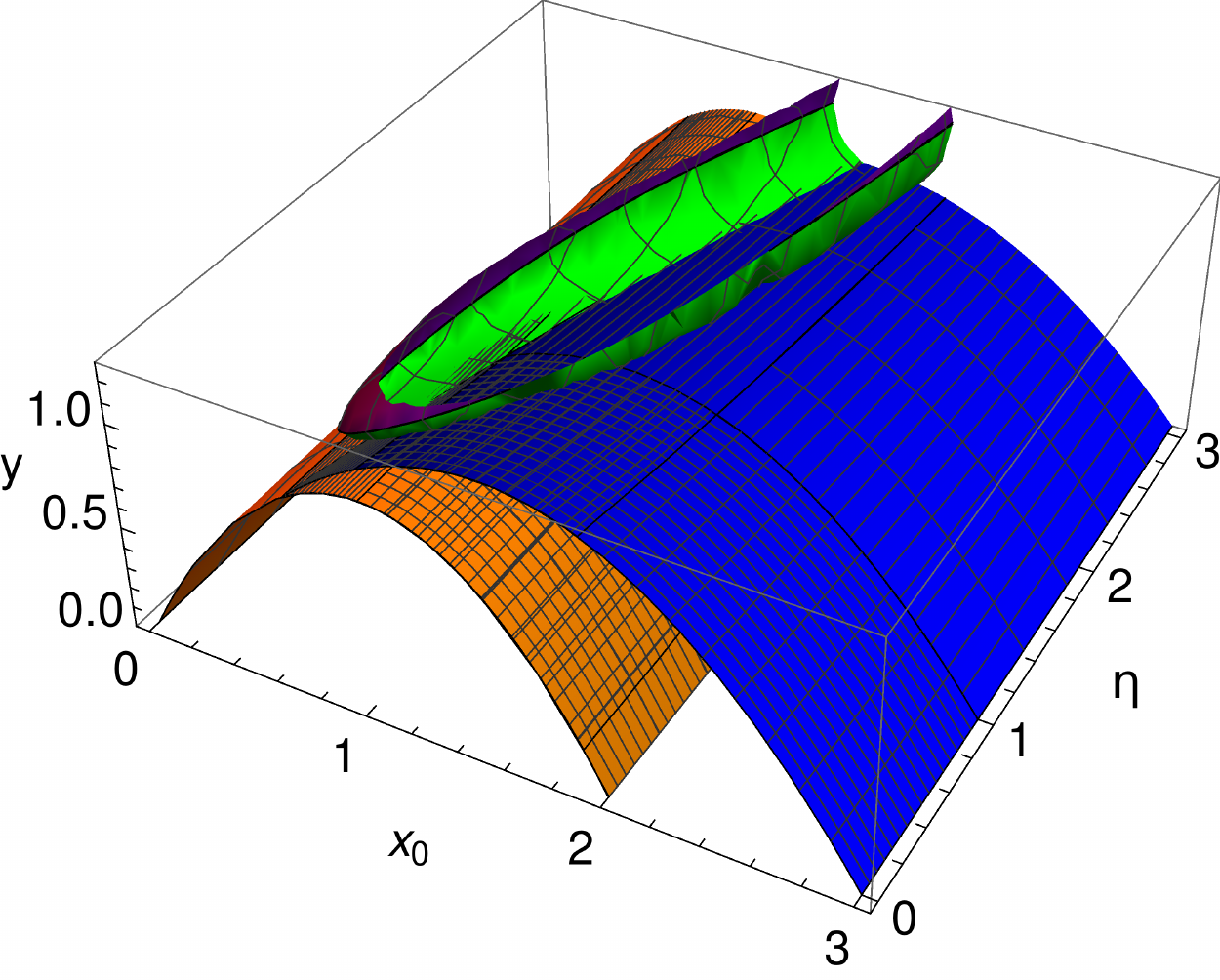}
	\caption{Three-dimensional plot of the limiting surfaces for $ \tilde{\gamma}_{_{RN}}>0$, $\omega_{_{RN}}^2 >0$ as a function of the dimensionless shell radius, $ x_0 $, the charge ratio, $ \eta $,  and the dimensionless charge, $ y$ (on $ \mathcal{M}_- $). The solutions with positive energy density lie between the orange and the blue surfaces, corresponding to $ y=x_0(2-x_0) $ and $y=  4 x_0(3- x_0)/(\eta +7) $, respectively (recall Eq.(\ref{eq:acrn})). Those configurations stable under perturbations are located either below the orange surface or inside the green and purple surface described by the functions of Eq.(\ref{omega RN2}). Remarkably, there is a non-vanishing overlapping region between the green and blue surfaces, described by Eq.(\ref{rel y-}), thus leading to stable solutions with positive energy density.}
	\label{fig:stabl}
\end{figure}

\begin{figure}[t!]
		\centering
	\includegraphics[width=0.44\textwidth]{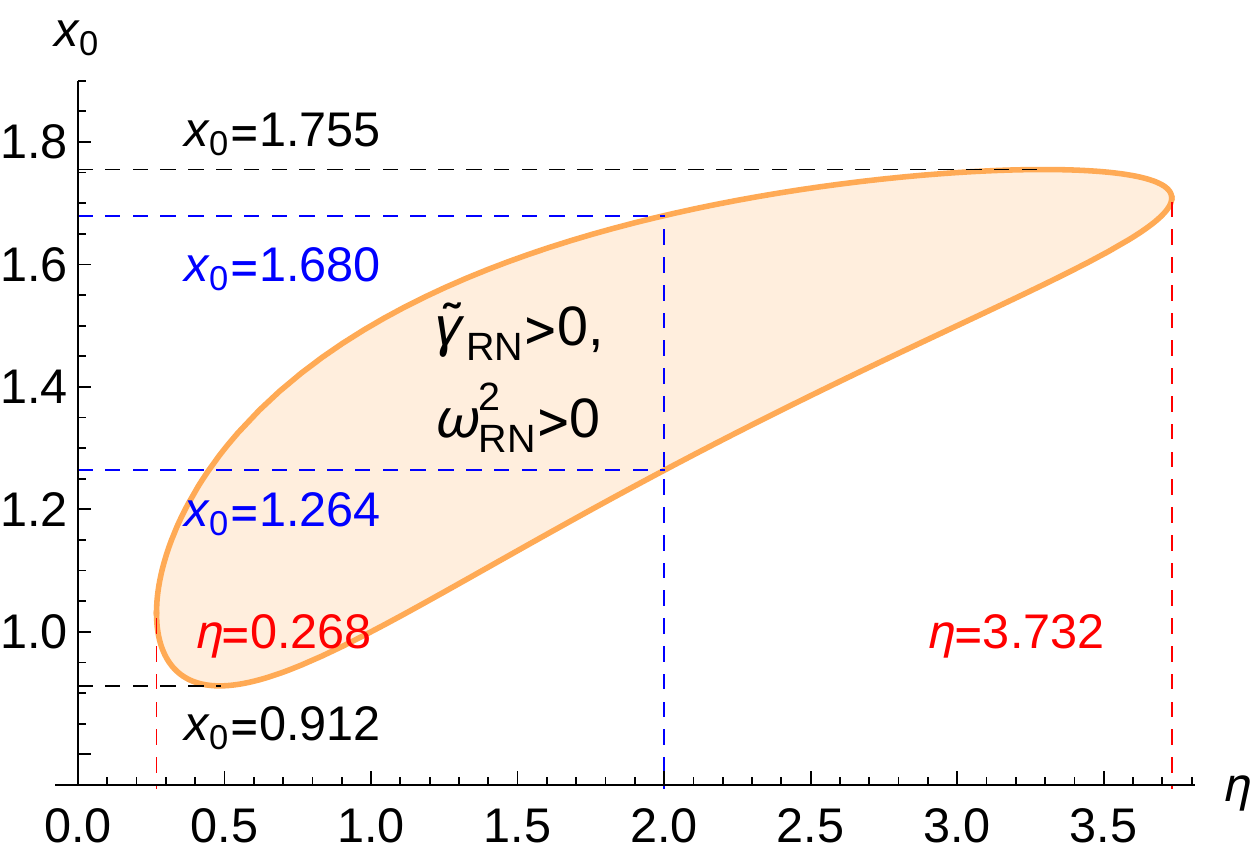}
	\caption{Parameter space for the region contained within (\ref{rel y-}) in terms of the charge ratio, $ \eta$, and of the dimensionless shell radius, $ x_0$. All those values falling inside the spaceship-shaped orange region will lead to stable, positive energy configurations. The blue dashed lines will be useful later when illustrating the double shadows.}
	\label{fig:etax0}
\end{figure}

Our next goal is to inspect the existence of overlapping regions between the positivity of the energy, $ \tilde{\gamma}_{_{RN}}>0$, and the stability of the corresponding solutions, $ \omega_{_{RN}}^2>0$. To this end, in Fig. \ref{fig:stabl} we depict the limiting surfaces for each of such features in terms of $x_0$ and $\eta$, which clearly shows the existence of such an overlapping region. The latter corresponds to the dimensionless charge being contained within the curves
\begin{equation}\label{rel y-}
 \dfrac{a-2\sqrt{b}}{\eta ^2-2 \eta +17}<y<\frac{4x_0(3-x_0)}{\eta +7} \ ,
\end{equation}
where the lower limit corresponds to the lower part of the $ \omega_{_{RN}}^2>0$ region (the green surface in Fig. \ref{fig:stabl}), whereas the upper limit is the upper function of $y$ when $ \tilde{\gamma}_{_{RN}}>0$ (the blue surface in Fig. \ref{fig:stabl}). The parameter space in terms of $ x_0 $ and $ \eta $ for which stable, positive-energy configurations are found is plotted in Fig. \ref{fig:etax0}. The corresponding values of the charge-to-mass ratio $y$ can be found by substituting these values of the parameter space in the above relations.

\begin{figure}[t!]
	\centering
	\includegraphics[width=0.35\textwidth]{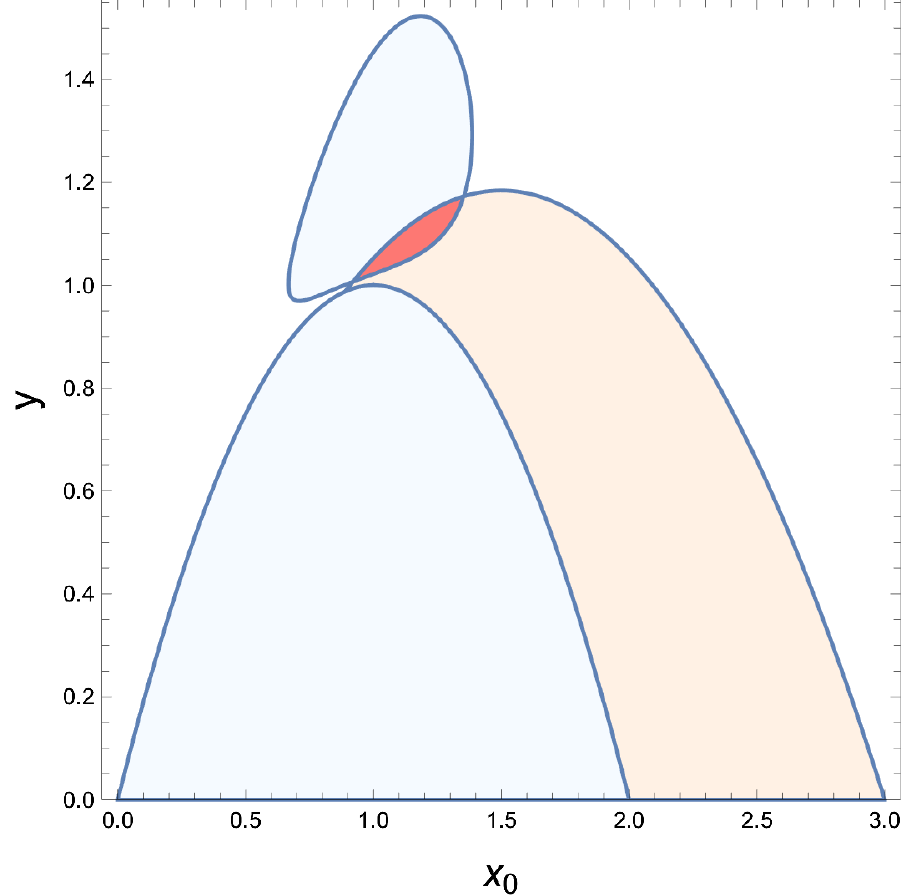}
	\caption{Representation of the domains with $\omega^2_{RN}>0$ (shaded light blue regions) and with $\tilde\gamma_{RN}>0$ (shaded orange) for a reference value $\eta=0.6$. Note the emergence of a new region with $\omega^2_{RN}>0$ above the parabola $y=x_0(2-x_0)$. This region always has bounded $\omega^2_{RN}$ everywhere (see Fig. \ref{fig:w2}), vanishing at the boundary. Note that now there is a non-empty intersection with the positive energy region (shaded red) which represents the parameter space with stable positive energy solutions.}
	\label{fig:eta0p6}
\end{figure}

In order to make the above explanation more transparent and easier to compare with the previous case using Fig. \ref{fig:stabilitysrn}, we cut different planes $ x_0-y $ at some $ \eta $ of Fig. \ref{fig:stabl}. Take for example $ \eta=0.6 $, for which the resulting plot is given in Fig. \ref{fig:eta0p6}, where the blue region corresponds to $ \omega_{_{RN}}^2>0 $ while the orange one to $ \tilde{\gamma}_{_{RN}}>0$. Now it is a piece of cake to see that there is a nonzero overlap between such regions (shaded in red) thanks to the new blue domain above the parabola $y=x_0(2-x_0)$ (compare with Fig. \ref{fig:stabilitysrn}) that emerges when $\eta\neq 0$ is not too big.

\begin{figure}[t!]
	\centering
	\includegraphics[width=0.44\textwidth]{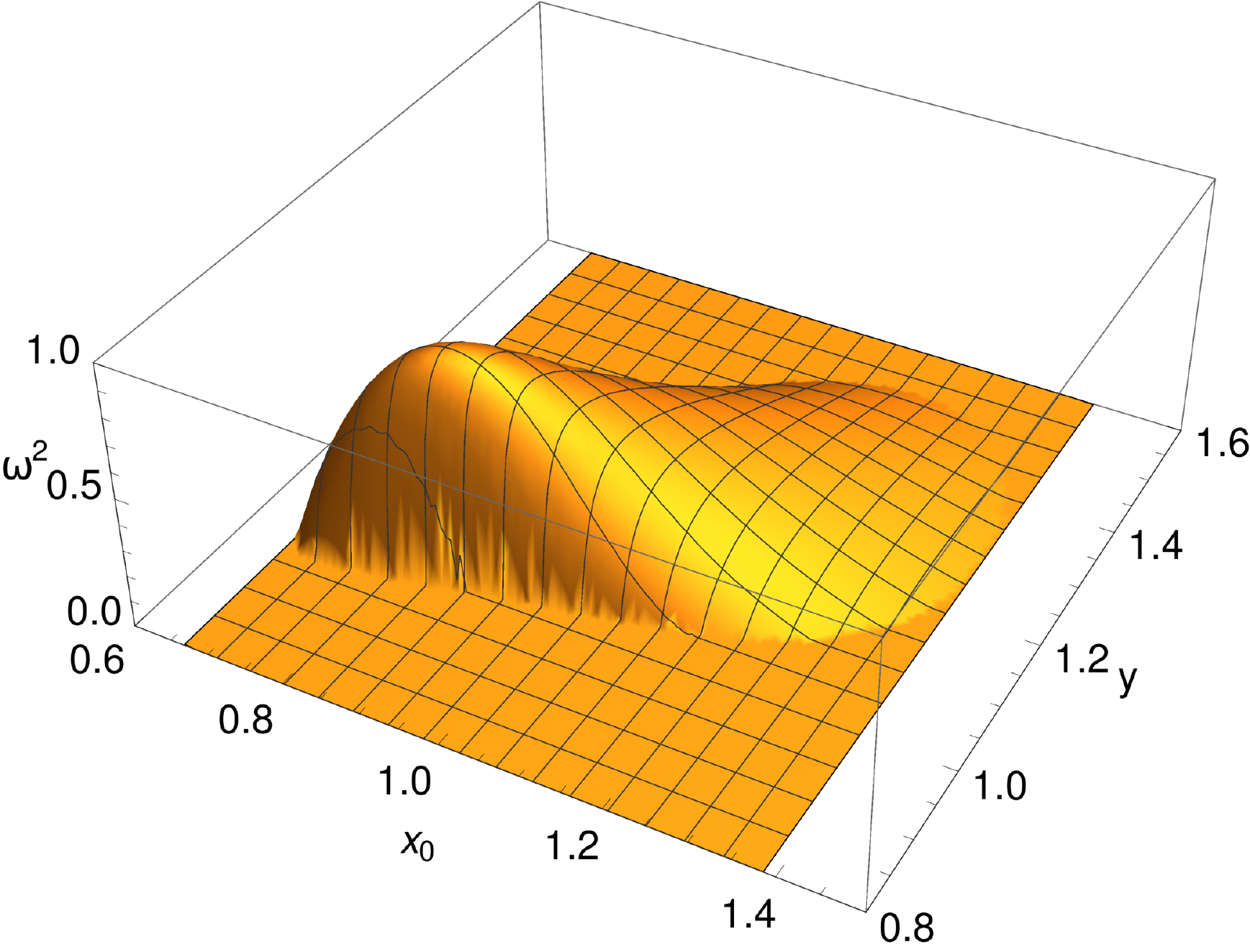}
	\caption{Representation of $\omega^2_{RN}>0$ on the $x_0-y$ plane for a reference value $\eta=0.6$. Note that $\omega^2_{RN}$ is always bounded and vanishes along the boundary of the blue shaded upper patch of Fig. \ref{fig:eta0p6}.}
	\label{fig:w2}
\end{figure}

\begin{figure}[h]
	\centering
	\includegraphics[width=0.44\textwidth]{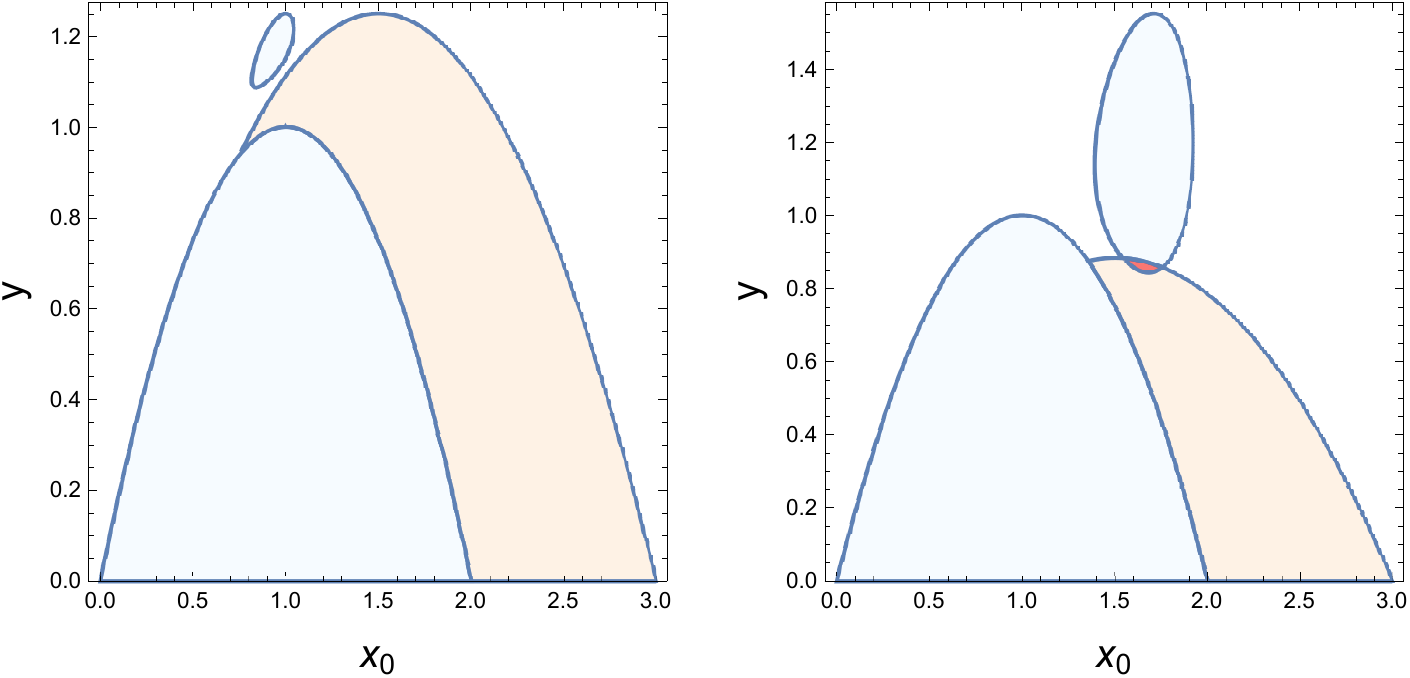}
	\caption{Same plots as in Fig. \ref{fig:eta0p6} but representing $\eta=0.2$ (left) and $\eta=3.2$ (right). Note how the overlapping red region does not exist for small values of $\eta$ and becomes smaller as $\eta$ grows, eventually disappearing  as well. The interval of $ \eta $ in which there exists an intersection is plotted in Fig. \ref{fig:etax0}. See also Figs. \ref{fig:all_eta} and \ref{fig:Physical_Domain} for more details on the evolution of the upper blue patch and the overlapping region as $\eta$ changes.}
	\label{fig:TwoEta}
\end{figure}

\begin{figure}[t!]
	\centering
	\includegraphics[width=0.35\textwidth]{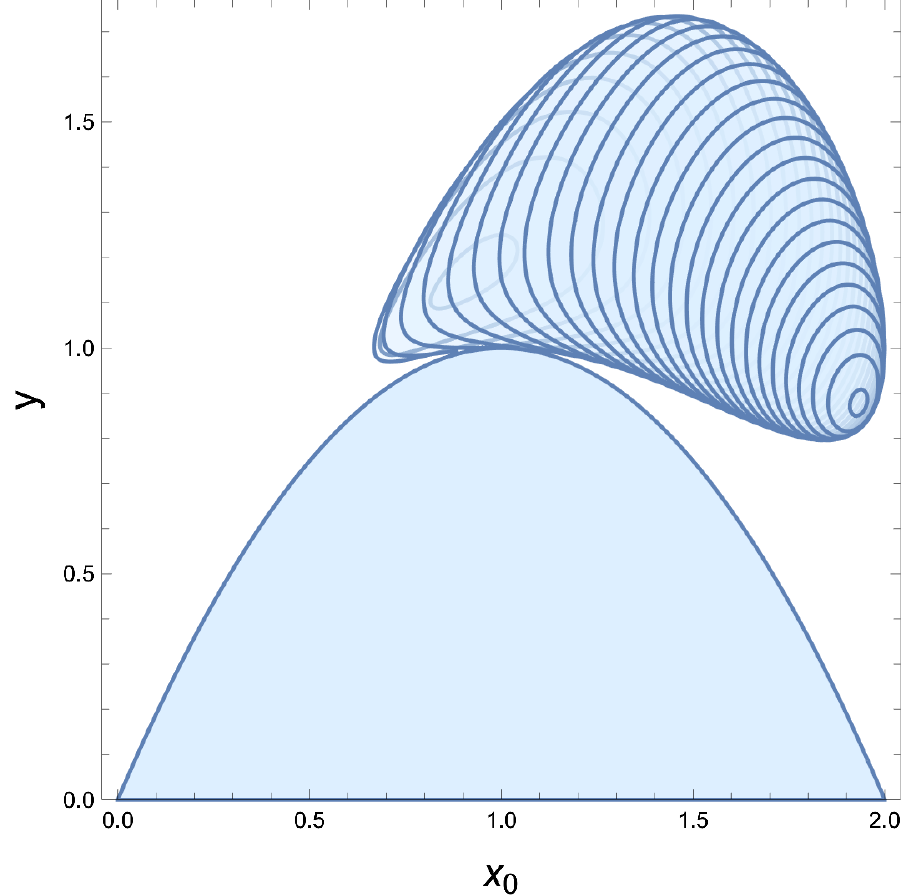}
	\caption{Representation of the regions with $\omega^2_{RN}>0$ as $\eta$ grows from zero to its maximum value $\eta\approx 5.85$. The contours drift from left to right as $\eta$ grows.}
	\label{fig:all_eta}
\end{figure}

\begin{figure}[t!]
	\centering
	\includegraphics[width=0.35\textwidth]{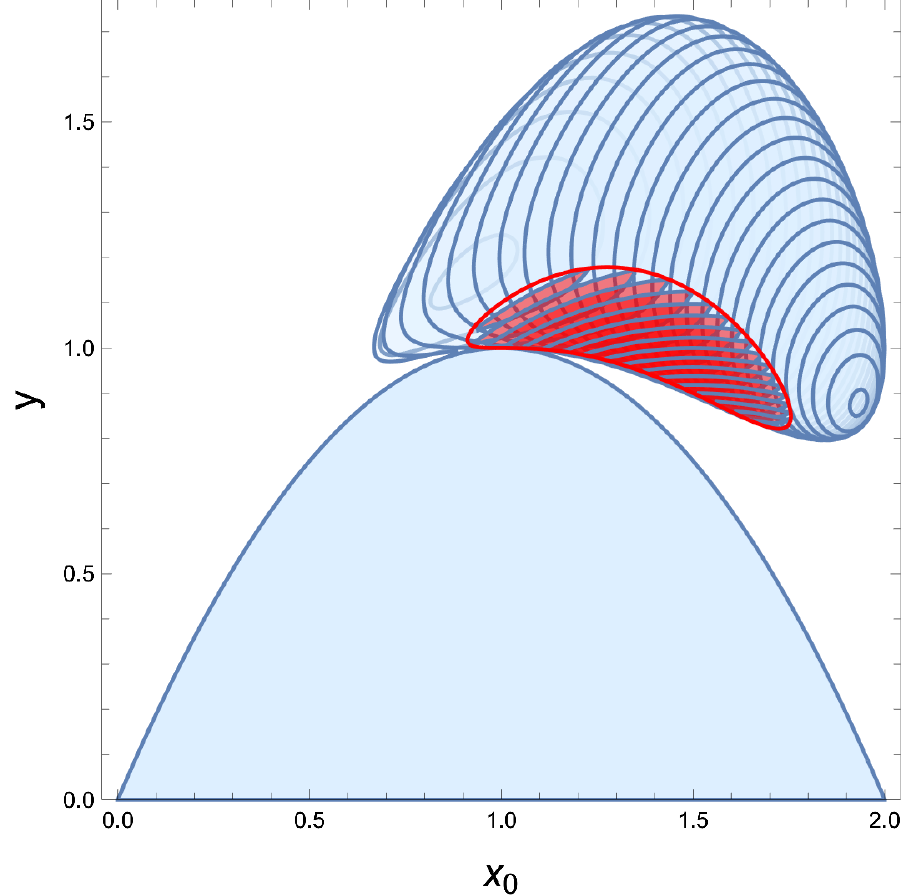}
	\caption{Representation of the regions with $\omega^2_{RN}>0$ for all values of $\eta$ including the overlap with $\tilde\gamma_{RN}>0$ (shaded red regions). The overlapping banana-shaped region admits an envelope curve, represented here in red color. That  patch corresponds to the projection on the $x-y$ plane of the (finite) volume of parameter space that represents stable, positive energy thin-shell solutions. }
	\label{fig:Physical_Domain}
\end{figure}

When considering different values of $\eta$ (see Fig. \ref{fig:TwoEta}) similar regions arise, the only difference being that the upper blue patch changes in size and shape and slightly moves from left to right as $\eta$ grows (see Fig. \ref{fig:all_eta}). Moreover, the orange region also changes with $\eta$ according to (\ref{eq:acrn}) and, therefore, the overlapping region sweeps as $\eta$ grows, as shown in Fig. \ref{fig:Physical_Domain}. The union of all the pink patches is bounded by an envelope curve shown in Fig. \ref{fig:Physical_Domain} as a red closed curve, which is exactly the same region whose projection on the $x-\eta$ plane is plotted in Fig. \ref{fig:etax0} but this time projected on the $ x_0-y $ plane. That bounded region represents the domain in parameter space that represents stable, positive energy thin-shell solutions.

\section{Structure of the asymmetric RN-RN wormhole: horizons and photon spheres}

We shall now look for the conditions of these stable, positive-energy density wormholes to have their shell radius  properly placed in order to represent traversable wormholes with a photon sphere on both sides of $\Sigma$. Regarding the horizon radius, since we are dealing with two RN space-times, their locations (on $\mathcal{M}_{\pm}$) are found by solving
\begin{equation}\label{horizon eq}
g^{rr}_{\pm}=0 \rightarrow r_h^2-2\,M_\pm r_h +Q^2_\pm =0 \ ,
\end{equation}
which leads to the following radius
\begin{equation}\label{horizons}
r_{h\pm} = M_\pm + (M_\pm^2 -Q_\pm^2)^{1/2} \ ,
\end{equation}
where we have kept the positive sign in front of the square-root because we are just interested here on the outermost (event) horizon. It is convenient to write such a radius in terms of the same dimensionless functions introduced in Eq.\eqref{R dim}-\eqref{Q dim} and \eqref{Q ratio}, which yields
\begin{eqnarray}
x_{h}^-&=& 1+\sqrt{1-y}\label{hor dim -} \\
x_{h}^+&=&\xi \,(1+\sqrt{1-\eta \,y/\xi^2})\label{hor dim +} \ ,
\end{eqnarray}
in $\mathcal{M}_-$ and $\mathcal{M}_+$, respectively. Therefore, event horizons are absent in $\mathcal{M}_-$ when $y>1$, and in $\mathcal{M}_+$ when $y>\xi^2/\eta$.

To compute the photon sphere, we note that it is defined as a region around a compact object with unstable circular null geodesics \cite{Claudel:2000yi}. From the point of view of an asymptotic observer it is determined by tracing back a given photon towards a bound orbit where it would have orbited the object many times before being released. The shape of this curve is therefore  entirely determined by the geometry of the compact object, and its interior is commonly known as the shadow\footnote{For a much more detailed discussion on photon spheres, black hole shadows, and lensing rings, see e.g. Ref. \cite{Gralla:2019xty}.}. Let us thus consider null geodesics, $ p^\mu p_\mu =0 $, which in a RN space-time read
\begin{equation}
-\frac{p_t^2}{f} +p_r^2 f+\frac{p_\phi^2}{r^2}=0 \ ,
\end{equation}
where $p^{\mu}=dx^{\mu}/ds$ is the photon four-momentum, $s$ is the affine parameter and $f=1-2M/r+Q^2/r^2$ the only non-vanishing metric function in the RN geometry, $ds^2=-fdt^2+f^{-1}dr^2+r^2d\Omega^2$ (on one of the sides of $\mathcal{M}_{\pm}$). Thanks to the staticity and spherical symmetry of the corresponding space-time,  both components $p_t$ and $p_{\phi}$ are conserved along the geodesic path. This allows to define the so-called impact parameter,  $ b= -p_\phi/p_t $, in such a way that the equation above can be rewritten as
\begin{equation}\label{mod geo}
\dfrac{1}{r^4}\left( \dfrac{dr}{d\phi}\right) ^2=\dfrac{1}{b^2}-V_{\text{eff}}(r) \ .
\end{equation}
This equation is akin to the motion of a particle in a one-dimensional effective potential of the form  $V_{\text{eff}}(r)= f(r)/r^2 $. Photons can propagate in those regions where the impact parameter satisfies $1/b^2-V_{eff} \geq 0$, with the equality representing the turning point. When such a point coincides with the maximum of the potential, then the light rays describe the last (unstable) orbit, defining the photon sphere of the object; this yields a  critical impact parameter $b_c=V_{max}^{-1/2}$. Thus, imposing the condition $ dV_{eff}/dr=0 $ on our two RN space-times, we find
\begin{equation}\label{photon phere}
r_{\gamma}^2-3M_{\pm}r_{\gamma} +2Q_\pm =0 \ .
\end{equation}
Using again our set of dimensionless variables, these two equations can be solved as
\begin{eqnarray}
x_{\g}^{-}&=&  \dfrac{3+ \sqrt{9-8 \,y}}{2} \label{photon sphere dim -} \\
x_{\g}^{+}&=& \xi \, \left( \dfrac{3+ \sqrt{9-8\eta \,y/\xi^2}}{2}\right) \ . \label{photon sphere dim +}
\end{eqnarray}

Our next task is to classify all possible combinations between the locations of the horizons and the photon spheres on each side.  Let us first study the manifold $ \mathcal{M}_- $, and combine the horizon equation (\ref{hor dim -}) with the photon sphere one (\ref{photon sphere dim -}). One then finds seven different possibilities:
 \begin{enumerate}
 	\item When $0 <x_0 \leq 1 $ and $0<y \leq 1$  then $ x_0 \leq x_{h}^{-} $, which means that the thin-shell wormhole will be always covered with a (event) horizon.
 	\item When $0 <x_0<1 $  and $y\in (1,9/8]$, then no horizon is present and the shell radius is always below the photon sphere.
 	\item When $1<x_0<3/2$ and there is no horizon ($y>1$), then a photon sphere is always present.
 	\item When $3/2<x_0<2$ and there is no horizon, then $y<\tfrac{x_0(3-x_0)}{2}$ must be fulfilled for a photon sphere to be present.
 	\item When $1 <x_0<2 $ and a horizon is present ($0<y \leq 1$) then $x_0(2-x_0)<y<1$ must be fulfilled, and a photon sphere is always present.
 	\item When $ 2 \leq x_0<3 $ then a horizon must be present and for a photon sphere to exist the condition  $0<y<\tfrac{x_0(3-x_0)}{2}$ must be fulfilled.
 	\item When $x_0 \geq 3$ no photon sphere can be present, as follows from a glance at Eq.\eqref{photon sphere rad-}, whose upper limit comes from setting $y \to 0$, yielding the photon sphere radius of the Schwarzschild solution, $x_0=3$.
 \end{enumerate}

We need now to add to this discussion the previously obtained parameter space for the existence of stable and positive-energy solutions, namely, Eq.(\ref{rel y-}), which is depicted in Fig. \ref{fig:etax0} in the plane $(\eta,x_0)$. As it can be seen, there is both a minimum and maximum absolute values for $x_0$, given by $x_0^{min} \approx 0.911 $ and $x_0^{max} \approx 1.75497$. In view of this constraint, from the above list it is clear that only the full case 3, and some parts of cases 2, 4, and 5  have the potential to represent traversable thin-shell wormholes having a photon sphere.

We now focus our attention upon the side $ \mathcal{M}_+ $ of the wormhole. In this case one might expect a similar discussion as in the $ \mathcal{M}_- $ case save by the replacement $y \to y \eta/\xi^2$ and a global factor $\xi$ in the scaling of the corresponding horizon and photon sphere radius (recall Eqs.(\ref{hor dim +}) and (\ref{photon sphere dim +})). However, enforcing the constraint (\ref{xi relation}), which allows to eliminate one of these variables in terms of the others and of the shell radius $x_0$, and by requiring that $x_h^+\leq x_0 \leq x_\gamma^+$, the complexity of classifying the different regions becomes far more noticeable than before. A subset of such regions is given by the following:
 \begin{itemize}
	\item $ 0<x_0\leq 1:$
	
	\subitem $0<\eta \leq \frac{x_0}{8-3 x_0}: x_0(2-x_0)\leq y\leq \frac{2x_0(3- x_0)}{\eta +3}$
	
\subitem	$ \frac{x_0}{8-3 x_0}<\eta <\frac{x_0}{2-x_0}: x_0(2-x_0)\leq y\leq y_L$
	
	\item $ 1<x_0<\frac{8}{5}:$
	
	\subitem $ 0<\eta \leq \frac{x_0}{8-3 x_0}: x_0(2-x_0)\leq y\leq \frac{2x_0(3- x_0)}{\eta +3}$
	
	\subitem  $ \frac{x_0}{8-3 x_0}<\eta <1: x_0(2-x_0)\leq y\leq y_L$
	
\subitem	$ 1<\eta <\frac{x_0}{2-x_0}: x_0(2-x_0)\leq y\leq y_L$
	
\item	$ \frac{8}{5}\leq x_0<2:$
	
\subitem	$0<\eta \leq \frac{x_0}{8-3 x_0}: x_0(2-x_0)\leq y\leq  \frac{2x_0(3- x_0)}{\eta +3}$
	
	\subitem $ \frac{x_0}{8-3 x_0}<\eta <1:  x_0(2-x_0)\leq y\leq y_L$
	
\subitem	$ 1<\eta <4: x_0(2-x_0)\leq y\leq y_L$
	
\item 	$ x_0=2: $
	
\subitem	$0<\eta <1: 0<y\leq \frac{4}{\eta +3}$
	
\subitem	$ 1<\eta <4: 0<y\leq \frac{4(\eta+1-2\eta^{1/2})}{(\eta-1)^2}$
	
\item 	$ 2<x_0<\frac{32}{13}: $
	
\subitem	$0<\eta \leq \frac{x_0}{8-3 x_0}: 0<y\leq \frac{2x_0(3- x_0)}{\eta +3}$
	
\subitem 	$ \frac{x_0}{8-3 x_0}<\eta <4: 0<y\leq y_L$
	
\item	$ \frac{32}{13}\leq x_0<3:$
	
\subitem	$ 0<\eta <4: 0<y\leq \frac{2x_0(3- x_0)}{\eta +3}$
\end{itemize}
where we have introduced the quantity
\begin{equation}\label{charge limit}
y_L=2\,\dfrac{\eta  \,x_0^2+(\eta-1 ) \, x_0- \sqrt{\eta ^2 \,x_0^4+2 \, (\eta-1)\,\eta  \,x_0^3}}{(\eta -1)^2} \ ,
\end{equation}
which is found by setting the squared root in Eq.\eqref{hor dim +} to zero and implementing the constraint (\ref{xi relation}); in other words, it is the maximum allowed charge in order to have a horizon in $ \mathcal{M}_+ $. It shall be also noticed that in some cases we may need to surpass this limit so we can find stable configurations hold by non-exotic matter sources. The maximum allowed value  is found by requiring to have a photon sphere, which comes from equalling the squared root in Eq.\eqref{photon sphere dim +} to zero and using again (\ref{xi relation}), and is given by the expression
\begin{equation}\label{y photon}
 y_p =\frac{2 \left(8 \eta  x_0^2-9 \eta  x_0+9 x_0\right)-8 \sqrt{4 \eta ^2 x_0^4-9 \eta ^2 x_0^3+9 \eta  x_0^3}}{9 (\eta -1)^2}.
\end{equation}
In addition, we point out that in some regions the lower limit of $ y $ is $ x_0(2-x_0) $, which coincides with the one needed to have $ \tilde{ \gamma}_{RN} >0$, as is written in Eq.\eqref{eq:acrn}. This might seem a pleasant coincidence but in fact a glance at Eq.\eqref{gammaRNRN} reveals that the denominator is the equation of the horizon (Eq.\eqref{horizon eq}) when $ R=R_0$, and after moving to dimensionless variables both equations become the same.  Consequently, when we set the horizon radius to be smaller or, at least, equal to the shell radius, this limit found in $ \tilde{ \gamma}_{RN} $ pops up.

\begin{figure}[t!]
	\centering
	\includegraphics[width=0.4\textwidth]{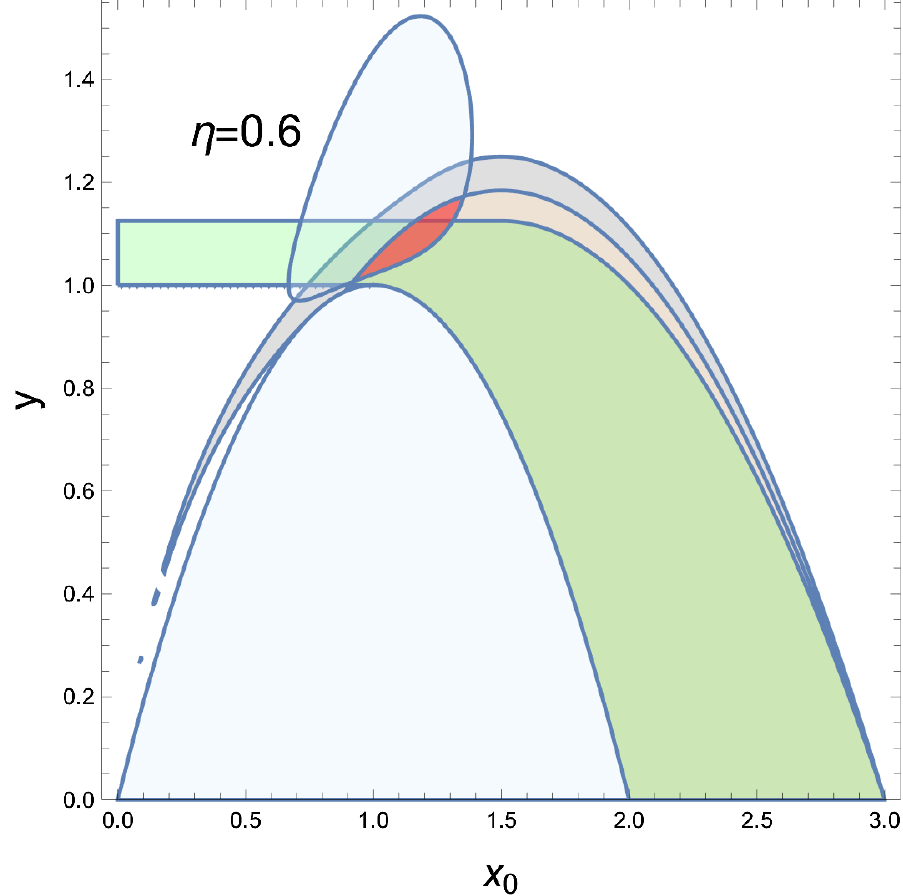}
	\caption{Representation of the parameter space regions that have a photon sphere in a traversable wormhole geometry on $\mathcal{M}^-$ (shaded green) and an analogous scenario on $\mathcal{M}^+$ (shaded gray) in the case $\eta=0.6$. Note that the shaded gray region has a large overlap with the green one and completely contains the orange region that represents $\tilde\gamma_{RN}>0$. Whenever gray and green overlap, we find a traversable wormhole with two photon spheres. When gray and green overlap on a blue region, those configurations are stable. If they meet on the red region, then they are stable and the thin shell has positive energy density.}
	\label{fig:Overlaps06}
\end{figure}

\begin{figure}[t!]
	\centering
	\includegraphics[width=8.5cm,height=4.5cm]{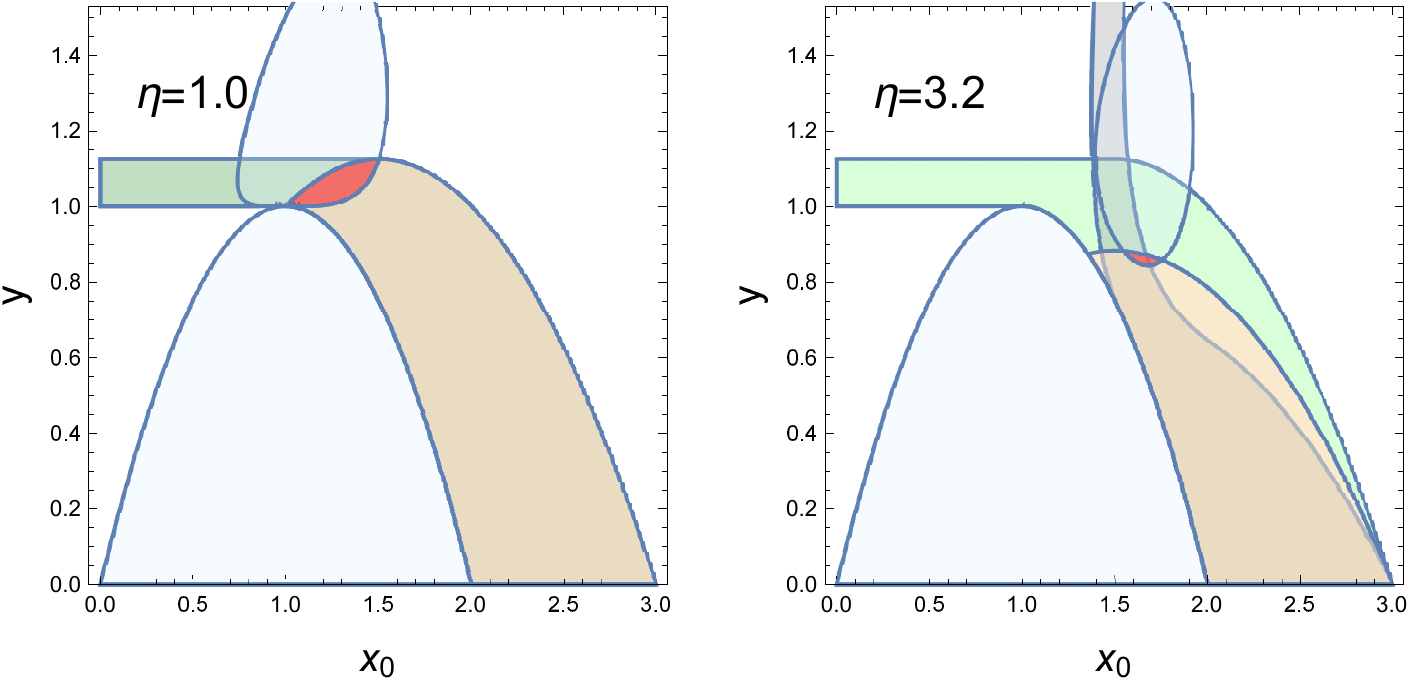}
	\caption{Same situation as in Fig. \ref{fig:Overlaps06} but with $\eta=1.0$ (left) and $\eta=3.2$ (right). When $\eta=1.0$, the gray and green shaded regions exactly coincide. Note how the upper blue patch representing positive frequencies  changes its shape and location with $\eta$, reducing its overlap with the orange region as $\eta$ grows. Though the green shaded region remains always the same, the gray one has significant  changes as $\eta$ grows. Remarkably, there exist large regions with two photon spheres, and one can always find stable regions with positive energy within this range of $\eta$. }
	\label{fig:TwoOverlaps}
\end{figure}

For a more visual discussion of the different regions in parameter space that lead to i) stable solutions, ii) stable and positive energy solutions, iii) stable solutions with one or two photon spheres, and iv) stable and positive energy solutions with one or two photon spheres, see Figs. \ref{fig:Overlaps06} and \ref{fig:TwoOverlaps} and their captions.

\begin{figure}[t!]
	\centering
	\includegraphics[width=0.45\textwidth]{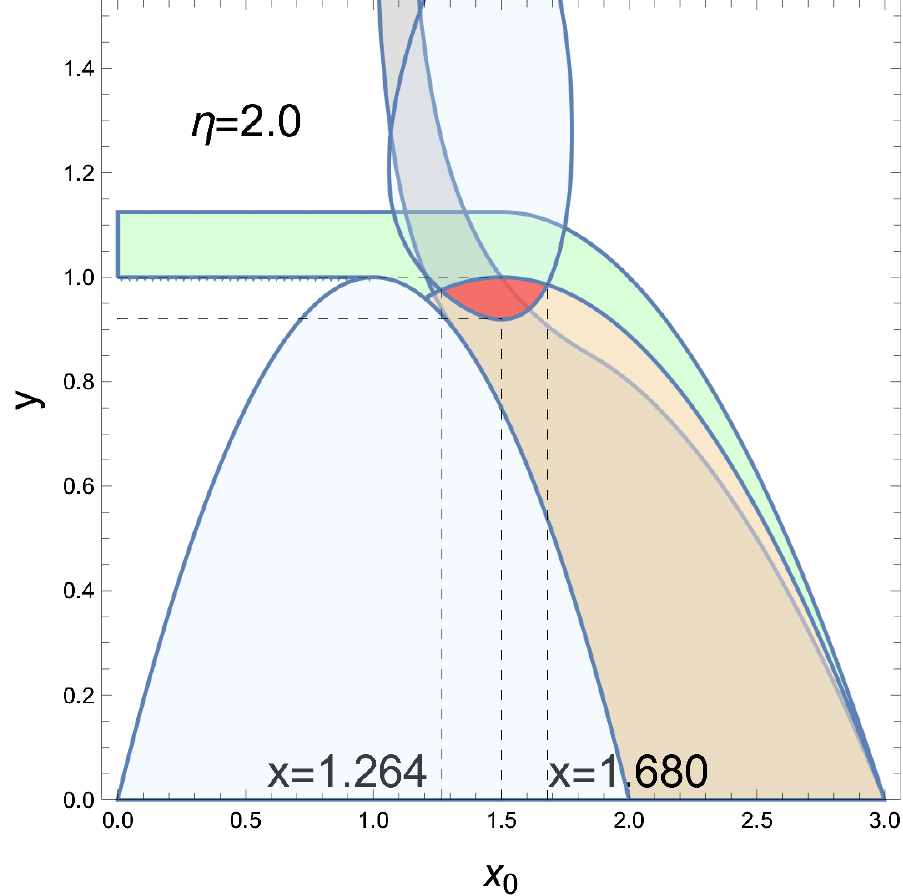}
	\caption{Same representation as in Figs. \ref{fig:Overlaps06} and \ref{fig:TwoOverlaps} but for $\eta=2.0$. The relevant points used in the discussion of the double shadow of Sec. \ref{sec:ds} are highlighted.}
	\label{fig:Overlaps20}
\end{figure}

\section{Double shadow} \label{sec:ds}

\begin{figure}[t!]
	\centering
	\includegraphics[width=8.5cm, height=5.5cm]{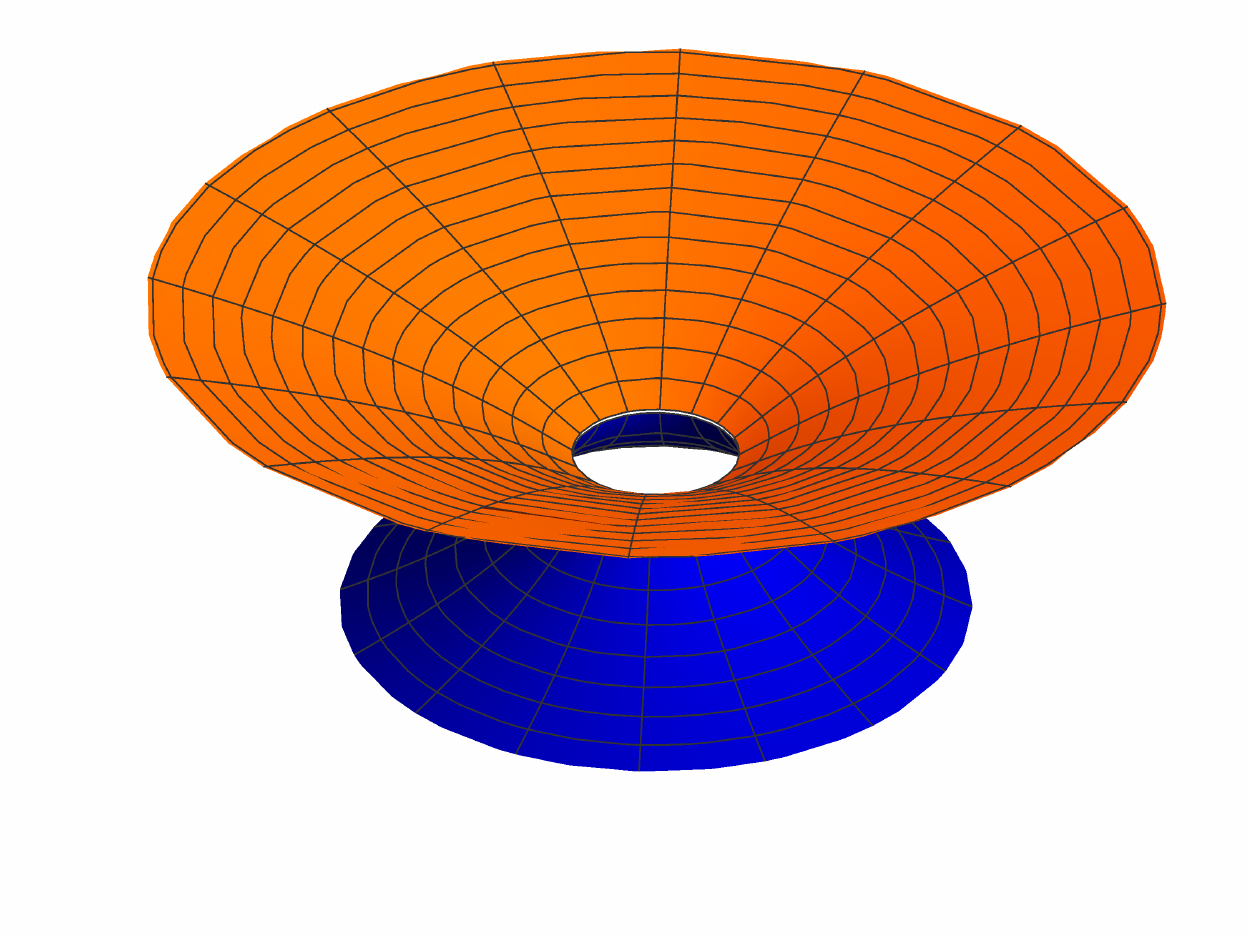}
	\caption{Euclidean embedding of the reflection-asymmetric RN-RN wormhole of Sec.\ref{sec:ds} on $\mathcal{M}_-$ (top orange) and $\mathcal{M}_+$ (bottom blue), corresponding to the parameters $ \eta =2 $, $ x_0=1.5 $ and $ y = 0.95 $,  which via the constraint (\ref{xi relation}) imply $\xi \approx 1.316$.}
	\label{fig:wh}
\end{figure}

Once we have identified the different regions that one may encounter in this reflection-asymmetric wormhole space-time, we can now focus on those stable, positive-energy configurations having a shell radius which lies above the event horizon  (when present)  but below the photon sphere (on each side) in order to characterize their shadows. For concreteness of our analysis, and given the large complexity of the spectrum of solutions depending on model parameters, we shall illustrate our results by setting the  charge-to-mass ratio to $ \eta=2 $ (see Fig. \ref{fig:Overlaps20}). For stable solutions with positive energy density within this choice, the corresponding dimensionless shell radius will have the lower and upper limits given by $1.264<x_0<1.680$ (see Fig. \ref{fig:eta0p6}). For this analysis we shall choose the value $ x_0=1.5 $. Plugging these values in Eq.\eqref{rel y-}, one finds the limits for the charge ratio $ 0.92 <y<1$ in order to have stable and positive-energy density solutions, which means we are in the case where the RN solution on $\mathcal{M}_-$ would have an event horizon (case 5 of our analysis above), but the shell radius is above it.  On the other hand, if  we compare this result with the upper limit to have an event horizon on $\mathcal{M}_+$ smaller than the shell radius $ x_0 $ then from Eq.(\ref{charge limit}) we find $ y_{L-}= 0.803848 $. This implies that  $ x_{h}^+ $ is imaginary, which means that there are no horizons on $\mathcal{M}_+$ regardless on where we place the shell. We choose then for convenience the value $ y= 0.95 $, which is clearly (see Fig. \ref{fig:Overlaps20}) within the desired (red) region of stability and positive-energy density. The last step is to use the constraint Eq.\eqref{xi relation} to determine the mass ratio for this case as $\xi \approx 1.316$. The reflection-asymmetric wormhole built this way is depicted in Fig. \ref{fig:wh}.

\begin{figure}[t!]
	\centering
	\includegraphics[width=8.5cm, height=5.5cm]{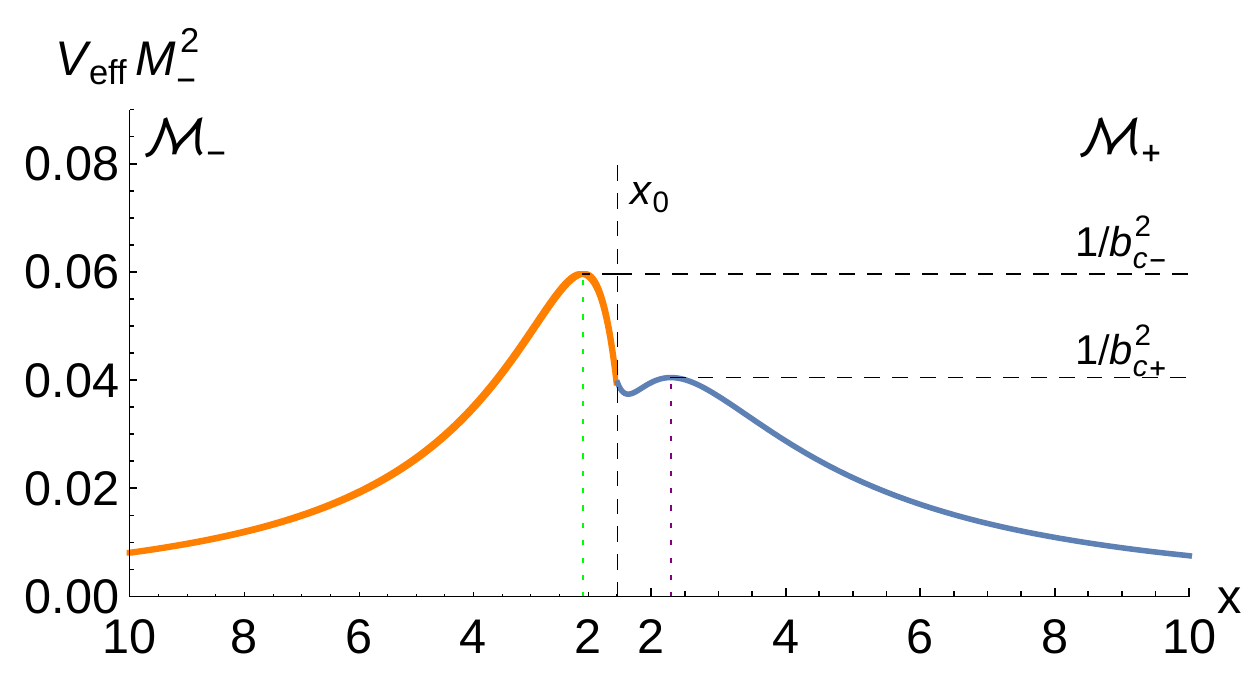}
	\caption{The effective potential $V_{eff}=f(r)/r^2$ on each side of the wormhole (blue  curve on $\mathcal{M}_+$ and orange on $\mathcal{M}_-)$ as a function of the dimensionless radius $x$ for the parameters $ \eta =2 $, $ x_0=1.5 $, $ y = 0.95 $ and  $\xi \approx 1.316$. The vertical black dashed line is the wormhole throat (the location of the shell), $x=x_0$, while the green and purple dotted vertical lines are the photon sphere radius of $ \mathcal{M}_- $ and $ \mathcal{M}_+ $, as given by (\ref{photon sphere rad-}) and (\ref{photon sphere rad+}), respectively. The horizontal black dashed lines are the maxima of the effective potential of each side and correspond to the critical impact parameters $1/b^2_{c\pm}$, as given by Eq.(\ref{critical impact parameter}) after its evaluation for the chosen values of this problem, yielding $b_c^+=4.969$ and $b_c^-=4.095$.}
	\label{fig:pot}
\end{figure}

Regarding the location of the  photon sphere radius, from Eqs.\eqref{photon sphere dim -} and \eqref{photon sphere dim +} we can compute their values for the choices above, which yields
\begin{eqnarray}
x_{\g}^-&=& 2.09161 \label{photon sphere rad-} \\
x_{\g}^+&=& 2.29221 \ , \label{photon sphere rad+}
\end{eqnarray}
which are larger than the shell radius $ x_0 $ on both sides, as expected. Recall that the effective potential $V_{eff}(r)$ attains a maximum at  these two radii, as shown in Fig. \ref{fig:pot}.

\begin{figure}[t!]
	\centering
	\includegraphics[width=0.44\textwidth]{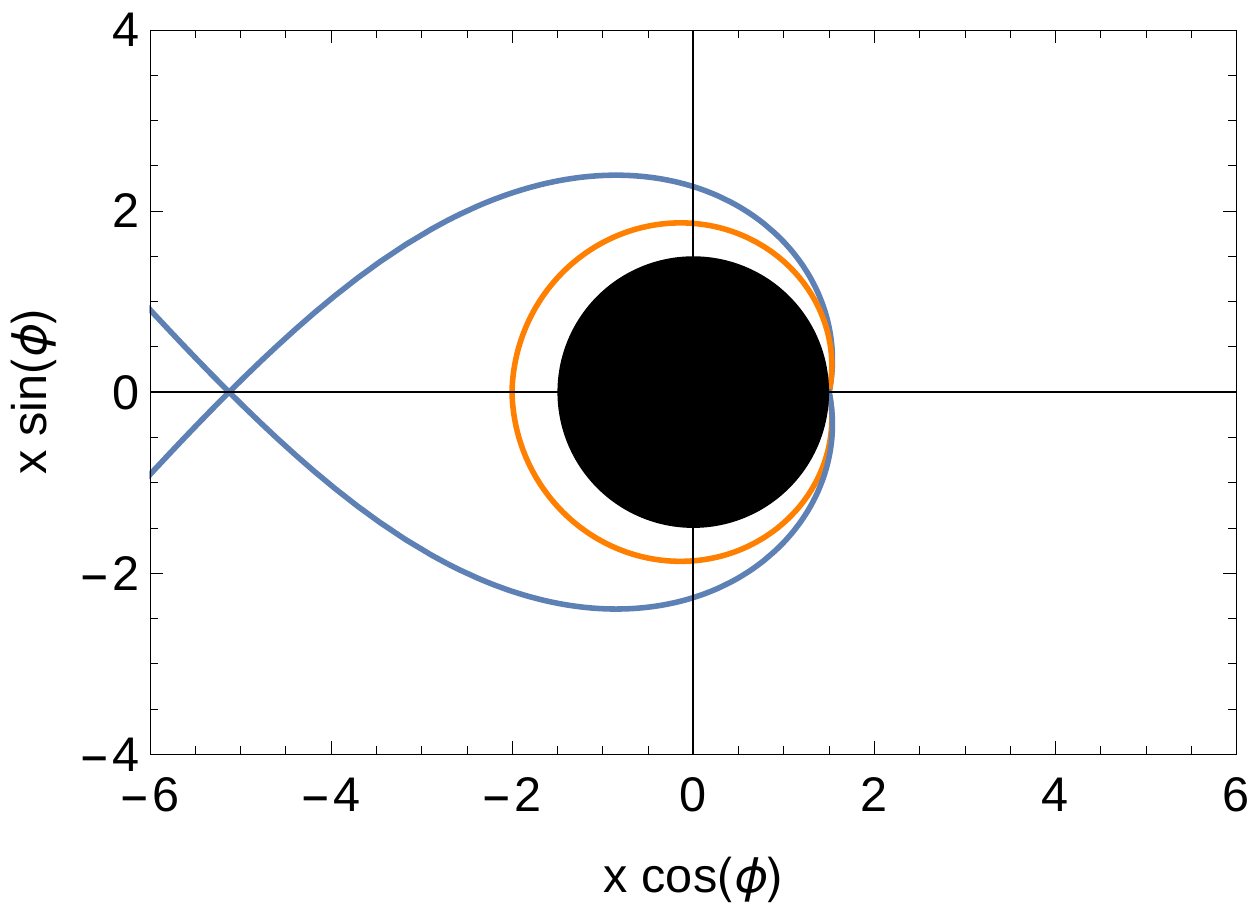}
	\caption{The light ray trajectory 
	when the impact parameter lies between the critical ones of each side of the wormhole, $4.095<\tilde{b}<4.969$, taking here the particular value $\tilde{b}=4.1$. The blue lines correspond to the light ray reaching and departing from the wormhole in $ \mathcal{M}_+ $, whereas the orange is the part of the trajectory in $ \mathcal{M}_- $. The black region corresponds to $x<x_0=1.5$ and represents the scissored part of the manifold, and does not belong to the available space-time.}
	\label{fig:geo}
\end{figure}

\begin{figure}[t!]
	\centering
	\includegraphics[width=0.44\textwidth]{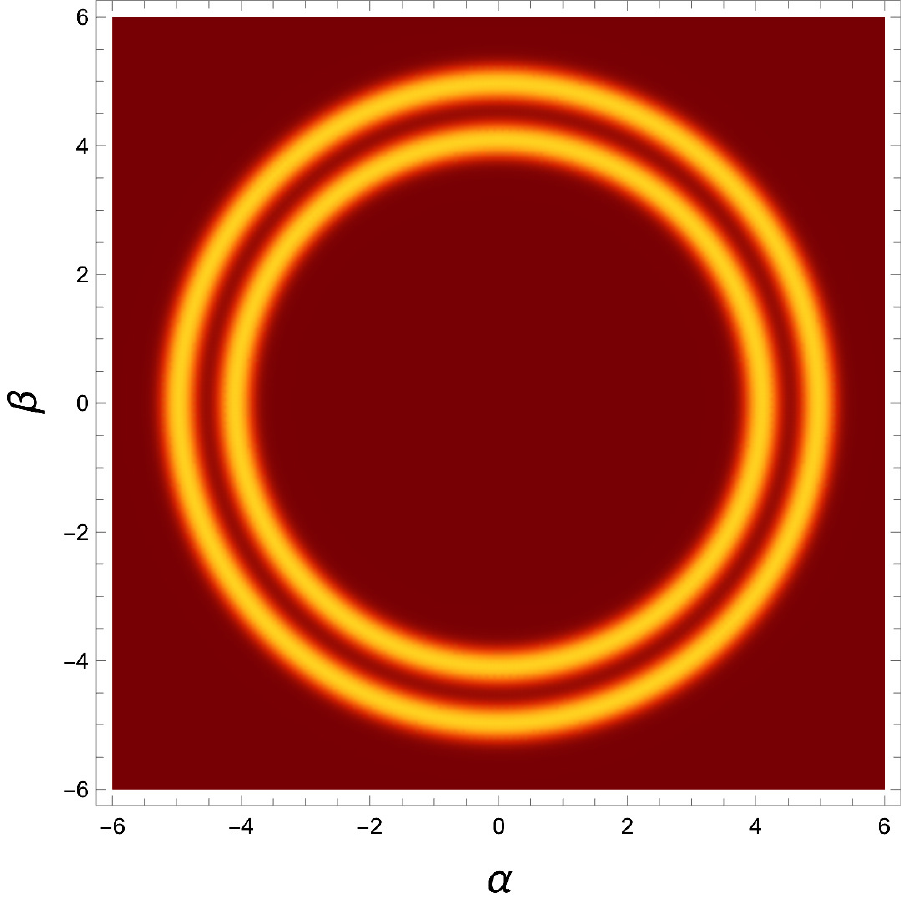}
	\caption{Illustration of the double shadow as seen
	 from the $ \mathcal{M}_+ $ side using celestial coordinates ($ \a $, $ \b $). From the $ \mathcal{M}_- $ region one would only see the inner ring. The dark region between the two rings represents the region of impact parameters contained between the critical values of  $ \mathcal{M}_+ $ and  $ \mathcal{M}_-$ (i.e. $ \tilde{b}_{c-}<\tilde{b_c}<\tilde{b}_{c+} $), whereas the dark region within the inner ring corresponds to impact parameters smaller than $\tilde{b}_{c-}$. We have used a gradient of colors to better illustrate the different intensities one would expect in the different regions. } 
	\label{fig:photon}
\end{figure}

With this setup we can analyze the critical parameter $ b_{c} $ leading to circular photon orbits of radius $ x_\gamma $ using Eq.(\ref{mod geo}), that corresponds to the value for which the right-hand side of this equation vanishes. Thus, rewriting this equation in terms of our dimensionless variables and evaluating at $ x_\gamma $ of each side, one finds that the impact parameter leading to unstable circular orbits (on each side) is given by
\begin{equation}\label{critical impact parameter}
\tilde{b}_{c\pm}= \dfrac{x_{\g\pm}}{\sqrt{f_{\pm}(x_\g)}} \ ,
\end{equation}
where $ \tilde{b} \equiv b/M_- $. As it can be seen from Fig. \ref{fig:pot}, for a light ray travelling from $\mathcal{M}_+$ to $\mathcal{M}_-$, when the impact parameter is larger than $ \tilde{b}_{c+}$ it will not be able to overcome the potential barrier and cross the throat, being deflected at a radius $x>x_{\gamma}^+$. However, when the impact parameter satisfies $ \tilde{b}_{c-}<\tilde{b_c}<\tilde{b}_{c+} $, which means we are in the region above the maximum of $V_{eff}^+$ but below the one of $V_{eff}^-$, then the light rays will pass through the wormhole throat and will bounce back to $\mathcal{M}_+$ due to the larger potential barrier of $ \mathcal{M}_- $. Therefore, observers on the manifold $ \mathcal{M}_+ $ will be able to see a double shadow (since there is no event horizon preventing the information to travel back), corresponding to the radii associated to the critical impact parameter on each side, while those observers on $\mathcal{M}_-$ will see the single shadow of its side, as usual. In Fig. \ref{fig:geo} we have plotted the light ray trajectory for the impact parameter $ \tilde{b}=4.1 $. Finally, by substituting the different values of the parameters into Eq.\eqref{critical impact parameter}, the corresponding photon rings can be plotted and an illustration of the effect is shown in Fig. \ref{fig:photon}, where the yellowish thick rings represent the regions with accumulation of photons that an observer from $\mathcal{M}_+$ would see.

\section{Conclusion}

In this work we have studied reflection-asymmetric thin-shell wormholes within Palatini $f(\mathcal{R})$ theories of gravity using a junction conditions formalism suitably adapted to the peculiarities of these theories. Such conditions involve a number of restrictions on the geometrical quantities and matter fields at the shell and across it, with the main highlight that, for spherically symmetric electrovacuum space-times, the number of effective degrees on the shell is reduced to just one, characterized solely by its energy density. Using these conditions we have built thin-shell wormholes from surgically joined Reissner-Nordstr\"om space-times on each side of the shell with different masses and charges. In the case in which one of the sides has a vanishing charge (surgically joined Schwarzschild-RN space-times) we have shown that it is not possible to have (linearly) stable solutions supported by positive energy density matter sources at the shell. However, in the full RN-RN thin-shell wormholes case a non-empty overlapping region with stable and positive-energy solutions is found, via some restrictions involving  the radius of the shell with the mass and charge of each side of the wormhole.

For such stable, positive-energy solutions we have carefully analyzed the conditions upon which the radius of the thin-shell is above the event horizon (when present) but below the photon sphere radius of each manifold $\mathcal{M}_{\pm}$. We found some regions where all these conditions (stability, positivity of the energy, traversability, and existence of  photon spheres) can be all met at the same time, graphically illustrated with some particular examples. This proves that within the Palatini $f(\mathcal{R})$ framework it is possible to find families of well behaved thin-shell wormhole geometries having two different photon spheres on each side, allowing for the generation of double shadows. Finally we have considered a particular example of model parameters, depicting its Euclidean embedding and the effective potentials on each side of the wormhole  as well as the corresponding light trajectories and double shadows. These results explicitly implement with a particular framework and specific solutions the detailed gravity-model-independent analysis carried out in Ref.\cite{Wielgus:2020uqz} on double shadows from reflection-asymmetric wormholes, and supports the feasibility of this approach with well behaved theories and solutions.

The relevance of these double shadows in the multimessenger astronomy era is promising: they open up a new avenue to find qualitatively new observational signals that may allow to discriminate black hole mimickers and exotic compact objects from the canonical ones in GR \cite{Ohgami:2015nra}, as well as to seek out traces of modified gravity. To this end, within the context of our framework one would need to generalize the toy model presented here to account for the presence of rotation in order to seek for multiple critical curves qualitatively different from the Kerr black hole expectations (for instance, similarly as done in Ref. \cite{Vincent:2020dij}). Though this is easier said than done due to the intrinsic complexity of the field equations of modified gravity, that largely prevents the finding of axially symmetric solutions of physical interest, recently a new procedure has been introduced within the context of Palatini theories of gravity which allows to find the counterpart of a GR-based solution on the modified gravity side using a set of algebraic transformations \cite{Afonso:2018bpv}. The power of this method has been recently exploited to find exact rotating black hole solutions within different such theories \cite{Guerrero:2020azx,Shao:2020weq}. We are currently working on implementing this strategy (analytically and numerically) to generalize the current results to more physically realistic scenarios, and hope to report soon on it.

\section*{Acknowledgements}
MG is funded by the predoctoral contract 2018-T1/TIC-10431.  DRG is funded by the \emph{Atracci\'on de Talento Investigador} programme of the Comunidad de Madrid (Spain) No. 2018-T1/TIC-10431, and acknowledges further support from the Ministerio de Ciencia, Innovaci\'on y Universidades (Spain) project No. PID2019-108485GB-I00/AEI/10.13039/501100011033, and the FCT projects No. PTDC/FIS-PAR/31938/2017 and PTDC/FIS-OUT/29048/2017.  This work is supported by the Spanish project  FIS2017-84440-C2-1-P (MINECO/FEDER, EU), the project PROMETEO/2020/079 (Generalitat Valenciana), and the Edital 006/2018 PRONEX (FAPESQ-PB/CNPQ, Brazil, Grant 0015/2019). This article is based upon work from COST Action CA18108, supported by COST (European Cooperation in Science and Technology).

\end{document}